\documentclass{ws-ijqi}
\usepackage{hyperref}
\usepackage[super,sort,compress]{cite}
\usepackage{graphicx,color}
\usepackage{dcolumn}
\usepackage{bm}
\usepackage{amsmath,amssymb,mathrsfs}
\usepackage{url}

\newcommand{\N}{\mathbb{N}}
\newcommand{\Z}{\mathbb{Z}}
\newcommand{\R}{\mathbb{R}}
\newcommand{\C}{\mathbb{C}}
\newcommand{\K}{\mathbb{K}}

\newcommand{\set}[1]{\mathsf{#1}}

\newcommand{\spc}[1]{\mathcal{#1}}

\def\d{{\rm d}}

\newcommand{\Span}{{\mathsf{Span}}}
\newcommand{\Lin}{\mathsf{Lin}}

\newcommand{\Herm}{{\mathsf{Herm}}}
\newcommand{\Supp}{{\mathsf{Supp}}}

\def\>{\rangle}
\def\<{\langle}

\newcommand{\map}[1]{\mathcal{#1}}
\newcommand{\Tr}{\operatorname{Tr}}

\newcommand{\Rank}{\mathsf{Rank}}

\newtheorem{theo}{Theorem}

\newtheorem{prop}{Proposition}
\newtheorem{cor}{Corollary}
\newtheorem{defi}{Definition}

\def\Proof{{\bf Proof.~}}
\def\qed{$\blacksquare$ \newline}

\newcounter{myctr}
\def\myitem{\refstepcounter{myctr}\bibfont\noindent\ifnum\themyctr>9\else\phantom{0}\fi\hangindent17pt\themyctr.\enskip}

\begin{document}

\catchline{}{}{}{}{}

\title{EXTREME QUANTUM STATES AND PROCESSES, AND EXTREME POINTS OF GENERAL SPECTRAHEDRA IN FINITE DIMENSIONAL ALGEBRAS}

\author{GIULIO CHIRIBELLA}

\address{QICI Quantum Information and Computation Initiative, Department of Computer Science, The University of Hong Kong, Pokfulam Road, Hong Kong\\
Hong Kong, China,\\
giulio@cs.hku.hk}

\address{Department of Computer Science, University of Oxford, Wolfson Building, 7 Parks Road,\\
 Oxford OX1 3QG, United Kingdom}
 
 \address{Perimeter Institute for Theoretical Physics, 31 Caroline St N, Waterloo, \\ON N2L 2Y5, Canada}

\maketitle

\begin{history}
\end{history}

\begin{abstract}
Convex sets of quantum states and  processes play a central role in quantum  theory and quantum information.    Many important examples of convex sets  in quantum theory are spectrahedra, that is, sets of positive operators subject to  affine constraints.   These examples include sets of quantum states with given expectation values of some observables of interest,  sets of multipartite quantum states with given marginals, sets of quantum measurements, channels,  and multitime quantum processes, as well as sets of higher-order quantum maps and quantum causal structures.   This contribution provides a characterisation of the extreme points of  general spectrahedra, and bounds on the ranks of the corresponding operators.   
The general results are applied to several special cases, and are used to retrieve classic results such as Choi's characterisation of  the extreme quantum channels, Parhasarathy's characterisation of the extreme quantum states with given marginals, and the quantum version of Birkhoff's theorem for qubit unital channels.  Finally, we propose a notion of  positive operator valued measures (POVMs) with general affine constraints for their normalisation, and we characterise the extremal POVMs.   
\end{abstract}




\section{Introduction}

 It is a great honour to contribute to this special issue celebrating Professor Alexander Holevo's eightieth birthday.  
  My initial exposure to his work came  through his  classic book  on the probabilistic and statistical aspects of quantum theory \cite{holevo2011probabilistic}. Some part of it, on symmetry groups and covariant measurements, 
  became the basis of my PhD  \cite{chiribella2004efficient,chiribella2004covariant,chiribella2004extremal,chiribella2005optimal,chiribella2006maximum,chiribella2006extremal,chiribella2006optimal}. Other parts, of more foundational nature,  had an impact on the longer term.   Among them there was the introductory chapter on statistical models, an early version of what is now called the framework of general probabilistic theories \cite{hardy2001quantum,barrett2007information,barnum2011information,chiribella2016quantum,dariano2017quantum}.

A recurrent theme through Holevo's book is the role of convex structures in physics and, more specifically, in quantum theory.    
  This theme came to play a major role in quantum information, where many questions can be cast as convex optimisation problems and semidefinite programs \cite{bengtsson2017geometry,heinosaari2011mathematical,chiribella2012optimal,chiribella2016optimal,watrous2018theory}.  A few paradigmatic examples include: optimising ensembles of quantum states for the transmission of information  \cite{holevo1973bounds,davies1978information,sasaki1999accessible,oreshkov2011optimal,dall2011informational,giovannetti2014ultimate}, distinguishing quantum states and processes \cite{holevo1973statistical,helstrom1976quantum,holevo1982testing,yuen1975optimum,eldar2001quantum,krovi2015optimal,buvzek1999optimalclocks}, optimising approximate quantum cloners \cite{werner1998optimal,keyl1999optimal,dariano2001optimal,dariano2003optimal,chiribella2005extremal} and other quantum machines  \cite{buvzek1999optimal,horodecki2002method,buscemi2003optimal}, studying quantum channels with memory  \cite{kretschmann2005quantum,gutoski2007toward,chiribella2008quantum,chiribella2008memory,pollock2018non,milz2021quantum} and  causal structures in quantum mechanics \cite{chiribella2009theoretical,oreshkov2012quantum,chiribella2013quantum,perinotti2017causal,bisio2019theoretical}.     In all these problems, the optimisation can be restricted without loss  of the extreme points of a suitable convex set.

   In this contribution I will offer some insight into the structure of convex sets that appear frequently in quantum information, including {\em e.g.} convex sets of quantum states with prescribed marginals, or prescribed  expectation values of certain observables, convex sets of quantum processes and quantum measurements satisfying different types 
   constraints.
   Mathematically, all these convex sets are  spectrahedra \cite{anjos2011handbook},   consisting of all positive operators satisfying a given set of linear constraints.  
    The extreme points of various spectrahedra arising in quantum theory have been characterised in a series of previous works \cite{holevo2011probabilistic,choi1975completely,parthasarathy1999extremal,landau1993birkhoff,ruskai2002analysis,parthasarathy2005extremal,rudolph2004extremal,winter2004extrinsic,dariano2004extremal,chiribella2004extremal,dariano2005classical,chiribella2005extremal,chiribella2006extremal,chiribella2007continuous,chiribella2010barycentric,farenick2011classical,dariano2011extremal,pellonpaa2012quantum,jenvcova2012extremality,jenvcova2013extremal,jenvcova2016convex}.  In the mathematical literature,  the extreme points of general spectrahedra have been studied  for full matrix algebras on the real field \cite{ramana1995some,pataki2000geometry} and in particular cases on the complex field  \cite{chua2016gram,netzer2023linear}.   The main result of this contribution is a unified characterisation of the extreme points of general spectrahedra in finite dimensional operator algebras. Our result includes as special cases Choi theorem on the characterisation of extreme quantum channels \cite{choi1975completely},  Parthasarathy's theorem on the characterisation of extreme bipartite states with fixed marginals \cite{parthasarathy2005extremal}, and the ``quantum Birkhoff theorem,'' stating that all bistochastic qubit channels can be decomposed as convex combinations of unitary channels \cite{landau1993birkhoff,mendl2009unital}.    In addition to retrieving existing results in a unified way, the main result of the paper also implies  new results on extreme quantum states and processes on real Hilbert spaces \cite{stueckelberg1960quantum,araki1980characterization},  a topic that have received renewed interest with the advent of quantum information \cite{wootters1990local,hardy2012limited,wootters2012entanglement,chiribella2010probabilistic,renou2021quantum}.    
 
Finally, we consider positive operator valued measures (POVMs) that are resolutions of operators in a given spectrahedron.   This notion includes as special cases the usual notion of POVMs as resolutions of the identity operator \cite{davies1970operational,holevo1973statistical,helstrom1976quantum,holevo2011probabilistic},  as well as other positive operator valued measures that can be used to describe  quantum instruments \cite{davies1970operational,ozawa1984quantum,holevo2003statistical}, sequential  measurement processes \cite{gutoski2007toward,chiribella2008quantum,chiribella2008memory,ziman2008process,chiribella2009theoretical} and quantum measurement processes with indefinite causal order  \cite{chiribella2019quantum,bavaresco2021strict}.
   Specifically, we consider operator valued measures  The main result  is the characterisation of the extremal generalised POVMs in finite dimensional systems, which includes as a special case a previous  result of Refs. \cite{chiribella2007continuous,chiribella2010barycentric} showing that the extremal   POVMs in finite dimensions have support on a finite set of outcomes.    
 
The remainder of the paper is structured as follows. Section \ref{sec:framework} reviews the notion of spectrahedron and provides several examples from quantum information theory.   Then, Section \ref{sec:main} provides the general characterisation of the extreme points and a bound on the rank of the corresponding operators.  Section \ref{sec:applications} illustrates applications of the general results to examples in quantum information. Finally, Section \ref{sec:POVM} analyses the structure of POVMs that are resolutions of elements of a given spectrahedron,  providing a characterisation of the extreme points in the case of finite dimensional operator algebras and compact Hausdorff spaces.

 \section{Spectrahedra}\label{sec:framework} 
 
In this section we present the main notations and definitions used in the rest of the paper.  

\subsection{Notation} 

Let $\spc H $ be a finite dimensional Hilbert space over a field $\K$, which can be either  the complex field $\C$ or the real field $\R$.  The dimension of $\spc H$ will be denoted by $d$ throughout the paper.   

We will often use the Dirac notation $|\psi\>  \in  \spc H$ for the elements of $\spc H$,   with $\<\psi|\varphi\>  \in  \K$ denoting the scalar product of $|\psi\>$ and $|\varphi\>$, and $\<\psi|  :   \spc H  \to  \K  $ denoting the linear functional that maps the vector  $|\varphi\>  \in \spc H$ into the scalar $\<\psi|\varphi\> \in  \K$.  For a $d$-dimensional Hilbert space $\spc H$, we will denote by $\{  |k\>\}_{k=1}^d \subset \spc H$ a fixed orthonormal basis.


Let $\Lin (\spc H)$ be the algebra of all linear operators from $\spc H$ to itself.  For an operator $A \in  L(\spc H)$, $A^\dag  \in  L(\spc  H)$ will denote the adjoint of $A$, that is, the operator uniquely determined by the relation $\<\varphi|   A^\dag  |\psi\>   =  \overline{\<\psi|  A|\varphi\>} \, , \forall |\varphi\>$.  When $\K=  \R$, the adjoint coincides with the transpose, denoted by $A^T$.  Let  $\Herm  (\spc H) \subseteq  \Lin (\spc H)$ the subset of Hermitian operators over $\spc H$, that is, the operators $A  \in  L(\spc H)$ such that $A^\dag  = A$.   The cone of positive  operators on  $\spc H$,denoted by $\Herm_+  (\spc H)$,  is the subset of Hermitian operators  $A\in \Herm (\spc H)$ satisfying the condition $\<\psi| A| \psi\>  \ge 0$  for every $|\psi\> \in  \spc H$. 
  

An affine subspace of   $L(\spc H)$ will be specified by a linear map  $\map L: \Lin (\spc H)\to \Lin  (\spc K)$, where $\spc K$ is a finite dimensional Hilbert space (either over $\R$ or over $\C$), and an operator  $B  \in  \Lin (\spc K)$, and consists of all operators $A  \in L(\spc H)$ satisfying the condition $\map L(A)  =  B$.

\subsection{Spectrahedra in $L(\spc H)$}

Spectrahedra  were originally defined  in a full matrix algebras over the real field \cite{ramana1995some,anjos2011handbook}. The following definition extends the notion to arbitrary operator algebras.

\begin{defi}[Spectrahedron in $L(\spc H)$]\label{def:spectra}  A {\em spectrahedron}   ${\sf S} \subset  L(\spc H)$  is a convex set of the form
\begin{align}\label{eq:spectra1}
{\sf S}:  =  \bigg\{ A  \in  \Herm_+  (\spc H)  ~\bigg|~ \map L_j   (A)  =  B_j  \qquad \forall  j\in  \{1,\dots, n\} \bigg\}\, ,
\end{align} 
where $\spc K_j$ is a Hilbert space,  $B_j \in  L(\spc K_j)$, and $\map L_j: \Lin (\spc H)\to \Lin  (\spc K_j)$ is linear map for every $\forall j\in \{1,\dots , n\}$. 
\end{defi}

\begin{defi}[Hermitian-defined spectrahedron]
We say that the spectrahedron $\sf S$ in Eq. (\ref{eq:spectra}) is {\em Hermitian-defined}  if all the linear maps $(\map L_j)_{j=1}^n$ are Hermitian-preserving, that is, if $\map L_j  (A)  \in  \Herm(\spc K_j)  \, ,\forall A  \in  \Herm (\spc H) \, ,\forall j\in  \{1,\dots,  n\}$. 
\end{defi}
Note that every spectrahedron $\sf S$  is a convex set, namely if the convex combination $p\,  A_1   +  (1-p) \,  A_2  \in  {\sf S}  \, , \forall  p\in  [0,1]$ whenever $A_1$ and $A_2$ belong to $\sf S$.   
In this paper we will restrict our attention to the case where all the Hilbert spaces $(\spc K_j)_{j=1}^n$ are finite dimensional.    Note also that the spectrahedron $\sf S$ defined in Eq.  (\ref{eq:spectra1}) is a closed set, and therefore it is compact whenever $\spc H$ is finite dimensional.  Hence, Krein-Millman's theorem \cite{krein1940on} guarantees that every element in $A \in  {\sf S}$ is can be approximated arbitrarily well by a convex combination of extreme points.    For completeness, we recall that an element $A \in  {\sf S}$ is an extreme point of $\sf S$ iff it cannot be written as a (non-trivial) convex combination, {\em i.e.}  iff the condition $A  =  p\,  A_1  +   (1-p) \, A_2$, for some $p\in  (0,1)$ and $A_1,  A_2 \in \sf S$ implies $A_1  =  A_2  =A$.

The simplest example of a Hermitian-defined spectrahedron is the set of all normalised quantum states on $\spc H$, consisting of the intersection between $\Herm_+  (\spc H)$ and the affine subspace $\{  A \in  \Lin  (\spc H)~|~ \Tr [A]   = 1\}$.  This example corresponds to  $n=1$,    $\spc K_1   =  \K$ (with $\K=  \R$ or $\K  = \C$),  $B_1   =  1$, and $\map L_1 (A)  :=  \Tr [A]$ in Definition \ref{def:spectra}.    

In the rest of this subsection we discuss a few other examples from quantum information theory.  All this examples happen to be examples of Hermitian-defined spectrahedra

\medskip  

{\em Quantum states with given  expectation values of a set of observables.}   Given a set of operators $O_j  \in  \Herm  (\spc H)$ and a set of real numbers $a_j$,  $j=1,\dots,  n-1$,  one can consider the spectrahedron consiting of all positive operators $\rho  \in \Herm_+  (\spc H)$ satisfying the conditions  
\begin{align}\label{givenexpectationvalues}
 \Tr  [  O_j   \rho ]   =  a_j  \, ,  \quad   \forall j\in  \{1,\dots,  n-1\}  \qquad{\rm and}  \qquad  \Tr [  \rho ]=1  
 \, .  
\end{align} 
Physically, this spectrahedron consists of all normalised quantum states with prescribed expectation values of the observables $\{  O_j\}_{j=1}^{n-1}$.

\medskip  

{\em Quantum states with given marginals.}   Suppose that  the Hilbert space $\spc H$ is of the tensor product form   $\spc H  =  \spc H_1 \otimes \spc H_2   \otimes \cdots \otimes \spc H_{k}$, for a finite family of Hilbert spaces  $(\spc H_{m})_{m=1}^{k}$.   In this case, one can consider the convex set of quantum states that have a prescribed set of marginals with respect to a given set of  (possibly overlapping)  subsystems  $\{   S_1,  \dots,    S_n\}$, with $S_j  \subseteq \{1,\dots,  k\}\, , \forall j\in  \{1,\dots, n\}$.  Characterising this convex set is known as the quantum marginal problem \cite{klyachko2004quantum,klyachko2006quantum,christandl2006spectra,schilling2015quantum}.    Mathematically, the set of quantum states with given marginals is the spectrahedron consisting of all operators $\rho  \in  \Herm_+   (\spc H)$ satisfying the conditions 
\begin{align}\label{givenmarginals}
 \Tr_{\overline S_j}  [  \rho]   =  \rho_j     \, ,  \quad   \forall j\in  \{1,\dots,  n\}  \, ,
\end{align}
where $\Tr_{\overline S_j}$ denotes the partial trace over all tensor factors $\spc H_m$ with $m\not \in S_j$, and each $\rho_j$ is a normalised density matrix on $\spc H_{S_j}: =  \bigotimes_{m\in S_j}  \,  \spc H_m$.   This set corresponds to the spectrahedron  with $\spc K_j  =  \spc H_{S_j}$,  $B_j  =  \rho_{S_j}$ and $\map L_j  (A)   =  \Tr_{\overline S_j}  [A]$  in  Definition \ref{def:spectra}.

  \medskip

 {\em Quantum channels, bistochastic channels, Gibbs-preserving channels, energy-preserving channels, and other completely positive maps subject to mapping constraints.}  Quantum channels \cite{holevo2003statistical,heinosaari2011mathematical} transforming an input system (with Hilbert space $\spc H_{\rm in}$) into an output system (with Hilbert space $\spc H_{\rm out}$) are described by completely positive linear maps  $ \map C:  \Lin  (\spc H_{\rm in}) \to \Lin  (\spc H_{\rm out})$, satisfying the condition  $\map C^\dag  (I_{\rm out})  =  I_{\rm in}$, where $\map C^\dag$ is the adjoint of $\map C$ with respect to the Hilbert-Schmidt product $\<  X,  Y\>  : =  \Tr[ X^\dag \, Y]$.

  The  Choi-Jamio\l kowski's isomorphism \cite{choi1975completely,jamiolkowski1972linear}  (see also \cite{bengtsson2017geometry})  implies that the set of linear maps from $ \Lin  (\spc H_{\rm in})$ to $\Lin  (\spc H_{\rm out})$ is in one-to-one correspondence with the set of bipartite operators  $L(\spc H_{\rm in} \otimes \spc H_{\rm out})$.   Specifically, Choi's version of the isomorphism sets up a one-to-one correspondence between the convex set of quantum channels and the spectrahedron 
   \begin{align}\label{channelconstraint}
  {\sf S}_{\rm chan}  : =  \Big\{ C  \in  \Herm_+ (\spc H_{\rm in} \otimes \spc H_{\rm out}) ~\Big|  ~  \Tr_{\rm out}  [  C]    =  I_{\rm in}  \Big\} \, ,
 \end{align} 
  where $\Tr_{\rm out}$ is the partial trace over $\spc H_{\rm out}$ and $I_{\rm in}$ is the identity operator on $\spc H_{\rm in}$.    
   This  spectrahedron  corresponds to the case  $n=1$,  $\spc K_1  =  \spc H_{\rm in}$,  $B_1  =  I_{\rm in}$ and $\map L_1  (A)   =  \Tr_{\rm out}  [A]$  in  Definition \ref{def:spectra}.
       
Another important set of maps is the set of  bistochastic quantum channels  \cite{landau1993birkhoff,mendl2009unital}, that is, quantum channels  satisfying the additional condition $\map C  (  I_{\rm in})  =  I_{\rm out}$.   Via Choi's representation,  the convex set of bistochastic channels is in one-to-one correspondence with the spectrahedron 
 \begin{align}\label{bistochasticchannels}
{\sf S}_{\rm bisto}  :  =\Big\{ C  \in  \Herm_+ (\spc H_{\rm in} \otimes \spc H_{\rm out})~\Big|~     \Tr_{\rm out}  [  C]    =  I_{\rm in} \, , ~ \Tr_{\rm in}  [  C]    =  I_{\rm out}\Big\}  \, .
 \end{align} 
  This spectrahedron coincides, up to a multiplicative factor, with the spectrahedron of  bipartite quantum states with maximally mixed marginals, which is a special case of the spectrahedra considered in the previous example.  
    
A generalisation of the notion of bistochastic channel is the notion of {\em Gibbs-preserving channel} \cite{faist2015gibbs},  that is, a quantum channel satisfying the additional condition $\map C  (\rho_{\rm in})  =  \rho_{\rm out}$, where $\rho_{\rm in} \in \Herm_+  (\spc H_{\rm in})$ and $\rho_{\rm out}  \in  \Herm  (\spc H_{\rm out})$ are two fixed states, which can be regarded as Gibbs states with respect to suitable Hamiltonians.   Via Choi's representation, the convex set of Gibbs-preserving channels is in one-to-one correspondence with the spectrahedron 

 \begin{align}\label{gibbspreas}
 {\sf S}_{\rm Gibbs}  :  =  \Big\{     C  \in  \Herm_+ (\spc H_{\rm in} \otimes \spc H_{\rm out}) ~\Big| ~\Tr_{\rm out}  [  C]    =  I_{\rm in}   \, , ~ \Tr_{\rm in}  \left[   ( \rho_{\rm in}^T \otimes I_{\rm out}) \,C\right]    =  \rho_{\rm out} \Big\} \, .
 \end{align}      

Another set with a similar structure is the set of quantum channels that preserve a set of observables, such as the energy  and  its momenta  \cite{chiribella2017optimal}.    In general, we can consider the set of channels $\map C$ that map a given set of observables $\{  B_j\}_{j=1}^{n-1}  \in  \Herm  (  \spc H_{\rm out})$ into another given set of observables $\{  A_j\}_{j=1}^{n-1}  \in  \Herm  (  \spc H_{\rm in})$, that is $\map C^\dag  (B_j)   =  A_j  \, , \forall j\in\{1,\dots, n\}$.    The channels in this set are in one-to-one correspondence with  operators $  C  \in  \Herm_+ (\spc H_{\rm in} \otimes \spc H_{\rm out}) $ satisfying the constraints
 \begin{align}\label{observablepres}
 \Tr_{\rm out}  [    (I_{\rm in} \otimes B_j  )  C]    =  A_j  \,  ,\qquad \forall j\in\{  1,\dots, n\}  \,, , 
 \end{align}      
 with  $A_n:  =I_{\rm in}$  and $B_n:  =  I_{\rm out}$.  
Sets of this form also include the set of all quantum channels that transform a given positive operator valued measure  (POVM) $ (Q_j)_{j=1}^n$, representing a measurement on the output system, into another given POVM $ (P_j)_{j=1}^n$, representing a measurement on the input system.  

All the above  sets can be treated in a unified way, by considering the convex set of completely positive linear maps subject to   {\em mapping constraints}, that is, constraints on how the maps  transform certain operators. 
Specifically, we consider completely positive maps $\map C:   \Lin  (\spc H_{\rm in}) \to \Lin  (\spc H_{\rm out})$ subject to the constraints
\begin{align}
\map C   (  A_j) =   B_j    \, ,  \forall j\in  \{1,\dots,  k\}    \qquad {\rm and}  \qquad \map C^\dag  (B_j) =  A_j  \, ,\forall j\in  \{k+1,\dots,  n\} \, ,    
\end{align}
for some $n,k  \in  \N$,  $A_j \in L (\spc H_{\rm in})$, and $B_j \in  L(\spc H_{\rm out})$,   $\forall j \in  \{1,\dots,  n\}$.   If all the operators $\{  A_j,  B_j\}_{j=1}^n$ are Hermitian, we say that the map $\map C$ is subject to {\em Hermitian mapping constraints}.  
Via Choi's representation, the above convex set is in one-to-one correspondence with the spectrahedron 

\begin{align}
\nonumber 
{\sf S}_{\rm map}     :  =  \Big\{   C  \in  \Herm_+ (\spc H_{\rm in} \otimes \spc H_{\rm out}) ~\Big | ~  & \Tr_{\rm in}  \left[   ( A_j^T \otimes I_{\rm out}) \,C\right]    =  B_j  \, , \forall j\in  \{1,\dots,  k\}\, ,  \\
&  \Tr_{\rm out}[  (  I_{\rm in}  \otimes B_j)\, C]   =    A_j  \, ,\forall j\in  \{k+1,\dots,  n\}   \Big\}   \, .   \label{mappingconstraints}
 \end{align}      
We call this spectrahedron the {\em spectrahedron of completely positive maps subject to Hermitian mapping constraints}.  
 Later in the paper, we will provide a characterisation of the extreme points of this spectrahedron, which will allow us to  treat quantum channels, bistochastic channels, Gibbs-preserving channels, and others in a unified way.  
  
\subsection{Spectrahedra in finite dimensional algebras}

The definition of spectrahedron can be extended  from the full operator algebra $\Lin  (\spc H)$ to arbitrary finite dimensional algebras.   This generalisation is useful in the study of quantum measurements, and hybrid quantum-classical systems.  

Let $\map A $ be a finite dimensional operator algebra over  a field $\K$, with either $\K  =  \R$ or $\K  = \C$. 
Let  $\spc A_{\Herm}  \subseteq \spc A$ be the real subspace consisting of all   Hermitian operators in $\spc A$, and let $\spc A_{\rm Herm,  +} \subset  \spc A_{\sf Herm}$ be the cone of all positive operators in $\spc A$.  We then have the following definition:  

\begin{defi}[Spectrahedron in a finite dimensional algebra]\label{def:spectraalg} 
 A {\em spectrahedron}   ${\sf S} \subset  \spc A$  is a convex set of the form
\begin{align}\label{eq:spectra}
{\sf S}:  =  \bigg\{ A  \in  \spc A_{\sf Herm ,+}    ~\bigg|~ \map L_j   (A)  =  B_j  \qquad \forall  j\in  \{1,\dots, n\} \bigg\}\, ,
\end{align} 
where $\spc K_j$ is a finite dimensional Hilbert space,  $B_j \in  L(\spc K_j)$, and $\map L_j: \spc A  \to \Lin  (\spc K_j)$ is linear map for every $j\in \{1,\dots , n\}$.
\end{defi}

\begin{defi}[Hermitian-defined spectrahedron in a finite dimensional algebra] 
We say that the spectrahedron ${\sf S} \subset  \spc A$ defined in Eq.  (\ref{eq:spectra}) is {\em Hermitian-defined} if all the linear maps $(\map L_j)_{j=1}^n$ are Hermitian-preserving,  that is, if $\map L_j  (A)  \in  \Herm(\spc K_j)  \, ,\forall A  \in  \spc A_{\rm \Herm} \, ,\forall j\in  \{1,\dots,  n\}$.    
\end{defi}

A few  examples of spectrahedra in finite dimensional algebras are presented in the following.  



\medskip  

{\em POVMs.}    The statistics of  quantum measurements is described by  positive operator valued measures (POVMs) \cite{davies1970operational,holevo1973statistical,helstrom1976quantum,holevo2011probabilistic}, defined as resolutions of the identity operator  that associate positive operators to possible events in a measurement process.  Here we will restrict our attention to  measurements with outcomes in a  finite set $\set X$,  for which a POVM is simply a collection of positive operators $  P_x\in  \Herm_+(\spc H) \,  , \forall x\in\set X$ satisfying the normalisation condition $\sum_{x \in \set X} P_x =  I$.     Equivalently, a finite-outcome POVM can be represented by a single operator  $\bigoplus_{x\in\set X}   P_x$  in the  algebra  $ \spc A  :  =  \bigoplus_{x\in\set X}  \,  L (  \spc H_x)$ with $\spc H_x \simeq \spc H$ for all $x\in\set X$.   With the notation of Definition \ref{def:spectraalg}, the convex set of all these POVMs  is the spectrahedron corresponding to $n=1$, $B_1=  I$, and $\map L_1\left( \bigoplus_{x\in\set X}  \,  A_x\right)  =   \sum_{x\in\set X}   A_x$.     

\medskip

{\em Quantum instruments.}  A quantum measurement process with outcomes in a finite set $\sf X$ is described by a quantum instrument \cite{davies1970operational,ozawa1984quantum,holevo2003statistical}, that is, a collection  of completely positive trace-non-increasing maps $\map M_x:     \Lin  (\spc H_{\rm in})  \to \Lin  (\spc H_{\rm out})$, $x\in \set X$, satisfying the condition that   $\sum_{x\in\sf X}  \map M_x$ is trace-preserving.   Through the Choi representation, the quantum instrument $(\map M_x)_{x\in \sf X}$  is associated to a family of positive operators $M_x  \in  \Herm_+  (  \spc H_{\rm in}\otimes \spc H_{\rm ou})$, which in turn can be  represented by the block-diagonal operator $M   :=  \bigoplus_{x\in \set X}   M_x$.       Hence, the convex set of quantum instruments with outcomes in $\set X$ is in one-to-one correspondence with a spectrahedron  in the algebra  $ \spc A: = \bigoplus_{x\in\sf X}  \,  \Lin  (\spc H_x)$, with $\spc H_x   \simeq  \spc H_{\rm in} \otimes \spc H_{\rm out} \, , \forall x\in\sf X$. 
 Specifically, the spectrahedron of quantum instruments consists  of all positive operators $M  \in  \spc A_{\Herm , +}$ satisfying the condition $\map L(M)  =  I$, where $\map L:  \spc A \to  L(\spc H_{\rm in})$ is the linear map defined by $\map L(  M)  = \sum_{  x\in\set X} \,   \Tr_{\rm out}  [  M_x]$.

\section{Characterisation of the extreme points}\label{sec:main}

In this section we provide the main results of the paper, which are two necessary and sufficient conditions for extremality in general spectrahedra, and  two bounds on the ranks of the extreme points. We will present all results in the general case where the spectrahedra are defined in arbitrary finite dimensional algebras.

\subsection{Extreme points of general spectrahedra}

A  first characterisation, along the lines of \cite{ramana1995some,pataki2000geometry},  is  as follows: 
\begin{theo}[Extreme points of general spectrahedra]\label{theo:extreme1}
Let  $\spc A $ be a finite dimensional algebra,  $\sf S   \subset  \spc A_{\Herm ,  +}$ be the spectrahedron   in  Eq. (\ref{eq:spectra}), and let  $\spc Z_{\sf S}  \subseteq \spc A_{\Herm}$   be the real subspace defined by  
\begin{align}
\spc Z_{\sf S}  : =  \big\{    H\in \spc A_{\Herm}   ~|~  \map L_j (H)   = 0  \, , \forall j\in\{1,\dots,  n\}  \big\} \, .
\end{align}  
   For an operator $A \in \sf S$, let $\spc V_A \subset  \spc A$ be the real subspace  defined by
\begin{align}\label{Arho}
\spc V_A  :  =  \big\{    H\in \spc A_{\Herm}   ~\big|~  \Supp(H)  \subseteq  \Supp  (A) \big\} \,  \,.
\end{align}  
The following are equivalent:  
\begin{enumerate}
\item the operator $A$ is an extreme point of $\sf S$, 
\item    the map  $\map M :  \spc  A  \to  \bigoplus_{j=1}^n  \Herm(\spc K_j)$ defined by 
\begin{align}\label{extremalitycondition1}  
\map M  (   H  ):  =  \bigoplus_{  j=1}^n    \,   \map L_i (H) \,  \qquad \forall H \in \spc A   
   \end{align}
to $ \spc V_A$  is injective on  $\spc V_{A}$,  namely $\forall H \in  \spc V_A:  ~ \map M (H) =  0 \Longrightarrow H =  0$,
\item the operators $  \big(  \map M(  H_m )\big)_{m=1}^{  \dim (\spc V_A) }$ are linearly independent whenever $\{   H_m\}_{m=1}^{\dim   (\spc V_A)}$ is a basis of $\spc V_A$,
\item  $\spc V_A \cup  \spc Z_{\sf S}  =  \{0\}$.  
\end{enumerate}
\end{theo}    
\Proof The proof adopts the method of perturbations.
    An operator $H  \in  \spc A$ is a perturbation of $A  \in \sf S$ iff there exists a positive number $\epsilon  >0$ such that  $A  +  t \,  H$ belongs to $\sf S$ for every $t\in  [ -\epsilon,  \epsilon]$.   The operator $A  \in  \sf S$ is an extreme point if and only if it  has only one  perturbation, namely the trivial one  $H =  0$.    

The condition that $H$ is a perturbation of $A$ can be made explicit by separately analyzing the conditions in Eq. (\ref{eq:spectra}). On the one hand,  $A+  t \,  H$ is positive for every $t\in  [ -\epsilon,  \epsilon]$ if and only if  $H$ is Hermitian and its support  is contained in the support of $A$, that is, if and only if  $H$ belongs to $\spc V_A$. On the other hand,    $\map L_j   (A  +  t \,  H)  =  B_j$  for every $t\in  [ -\epsilon,  \epsilon]$ if and only if $\map L_j  (H  )=0$. This condition holds for every $j\in  \{1,\dots, n\}$ if and only if $\map M (H)  =  0$, or equivalently, if and only if   $H   \in  \spc Z_{\sf S}$.   In summary, $H$ is a perturbation if and only if $H \in  \spc V_A$ and $\map M (H)  = 0$, or equivalently, if and only if   $H  \in  \spc V_A$ and $H   \in  \spc Z_{\sf S}$.  Extremality of $A$ is then equivalent to the injectivity condition  $\forall H \in  \spc V_A:  ~ \map M (H) =  0 \Longrightarrow H =  0$ and to the condition that $\spc V_A \cup    \spc Z_{\sf S}= \{0\}$.   This proves the equivalence of Condition 1 with Conditions 2 and 3.   Finally, Conditions 2 and 3 are obviously equivalent.   \qed  

\medskip  
Theorem \ref{theo:extreme1} implies the following corollary:  
\begin{cor}
If $A $ is an extreme point of the spectrahedron in Eq.  (\ref{eq:spectra}), then 
\begin{align}
\nonumber 
\dim (\spc V_A) &  \le   \Rank (\map M)  \le  \sum_{j=1}^n    \Rank   (\map L_j)  \le    \sum_{j=1}^n   \min \Big\{\dim  (\spc A) ,  \dim (\Herm  (\spc K_j)) \Big\}\\
  &  \le  \sum_{j=1}^n     \dim (\Herm  (\spc K_j))  \, ,   \label{bound}
\end{align}
and 
\begin{align}\label{pataki}
\dim (\spc V_A)  \le \dim (\spc A_{\Herm})    -   \dim (\spc Z_{\sf S}) \, .  
\end{align}
where $\dim  (V)$ denotes the dimension of an arbitrary  vector space $V$ over the real field, and $\Rank(\map L)$ denotes the rank of an arbitrary linear map $\map L:  V\to W$ between two vector spaces $V$ and $W$. 
\end{cor}
\Proof Eq. (\ref{bound}) is immediate from  Eq.  (\ref{extremalitycondition1}).  Eq. (\ref{pataki}) follows from Condition 3 in Theorem \ref{theo:extreme1}, which implies the bound  
\begin{align}
\nonumber 0   &=    \dim   (  \spc V_A  \cup   \spc Z_{\sf S} )  \\
\nonumber    &  = \dim (\spc A_{\rm Herm})   -       \dim   (\spc V_A  \cup   \spc Z_{\sf S} )^\perp\\
\nonumber &  \le     \dim (\spc A_{\rm Herm})   -       \dim   ( \spc V_A)^\perp  -  \dim  (    \spc Z_{\sf S} )^\perp\\
&=     \dim   ( \spc V_A)  +    \dim  (    \spc Z_{\sf S} )  -   \dim (\spc A_{\rm Herm}) \,.
\end{align} 
\qed  

In the special case where $\spc A $ is the full matrix algebra $M_d (\R)$, Eq. (\ref{pataki}) reduces to Pataki's inequality $r(r+1)/2  \le   d(d+1)/2   -   \dim (\spc Z_{\sf S})$, where $r$ is the rank of $A$ \cite{pataki2000geometry}.   When $\spc A$ is the full matrix algebra $M_d (\C)$, it reduces to the inequality $r^2  \le   d^2   -   \dim (\spc Z_{\sf S})$, proven by   Chua,  Plaumann, Sinn, and Vinzant \cite{chua2016gram}.

In general, the bounds (\ref{bound})  and (\ref{pataki})  can be made more explicit by  using the fact that every finite dimensional  algebra $\spc A$ is of the direct sum form
\begin{align}\label{algebradecomp}
\spc A   =  \bigoplus_{x \in \set X}  \,   L(  \spc R_x)  \otimes    I_{\spc M_x}  \, ,
\end{align}
where $\set X$ is a finite set, $\spc R_x$  and $\spc M_x$ are  finite dimensional Hilbert spaces, and $I_{\spc M_x}$ is the identity operator on $\spc M_x$.

Eq.  (\ref{algebradecomp}) implies that every positive operator $A \in  \spc A_{\Herm ,  +}$ can be decomposed as  
\begin{align}\label{rhodecomp}
A =  \bigoplus_{x \in \set X}    \, A_x \otimes I_{\spc M_x} \qquad {\rm with}  \qquad A_x\in  \Herm_+   (\spc R_x)  \, , \forall x\in\set X \, ,
\end{align}
and therefore  
\begin{align}\label{arhoexp}
\spc V_A   =    \bigoplus_{x\in\set X}   \,  \Herm(\Supp (A_x))  \otimes  I_{\spc M_x} \, , 
\end{align}
where $\Supp (A_x)  \subseteq \spc R_x$ denotes the support of $A_x$.  

Using Eq. (\ref{arhoexp}), we then obtain the following corollary:
\begin{cor}[Bounds on the ranks]
Let $A $ be an extreme point of the spectrahedron  $\sf S$ defined in Eq.  (\ref{eq:spectra}), and let   $r_x$ be the rank of the operator $A_x$ in Eq.  (\ref{rhodecomp}).
If   $\spc A$ is an algebra over $\C$ and each $(\spc K_j)_{j=1}^n$ is a complex Hilbert space, then 
\begin{align}\label{boundC}
\sum_{x  \in \set X}   r_x^2   \le  \sum_{j=1}^n    d_j^2  \, , 
\end{align}
where   $d_j$ is the dimension of $\spc K_j$, for every $j\in  \{1,\dots,  n\}$.  If   $\spc A$ is an algebra over $\R$ and  each $(\spc K_j)_{j=1}^n$ is a real Hilbert space,  then 
\begin{align}\label{boundR}
\sum_{x  \in \set X}   r_x (r_x+1)   \le  \sum_{j=1}^n    d_j(d_j+1)  \, . 
\end{align}
\end{cor}
\Proof  Immediate from Eq. (\ref{bound}) and from the fact that the dimension of the subspace of Hermitian operators on a $d$-dimensional Hilbert space  is $d^2$ in the complex case and $d(d+1)/2$ in the real case. \qed  

In Section  (\ref{sec:applications}) we will show several applications of the bounds (\ref{boundC}) and (\ref{boundR}).  

\subsection{Extreme points of Hermitian-defined spectrahedra}

When the spectrahedron is Hermitian-defined, a more elaborate characterisation becomes possible:  

\begin{theo}[Extreme points of Hermitian-defined spectrahedra]\label{theo:extreme}
Let   $\spc A $ be a finite dimensional algebra   and let  $\sf S \subset \spc A_{\Herm,  + }$ be the Hermitian-defined spectrahedron in Eq. (\ref{eq:spectra}).    An operator $A  \in \sf S$ 
 is an extreme point of $\sf S$ if and only if    
\begin{align}\label{extremalitycondition}  
\spc V_A   =   \Span_\R  \Big\{      \Pi_A \,  \map L_j^\dag   \big(  \Herm  (\spc K_j) \big)   \,  \Pi_A  \Big\}_{j=1}^n \, ,
   \end{align}
   where  $\Pi_A \in \spc V_A$ is the projector onto the support of $A$,   $\map L_j^\dag  :  L(\spc K_j) \mapsto  \spc A$ is the Hilbert-Schmidt adjoint of the map $\map L_j$, uniquely defined by the relation $\Tr [  X^\dag  \map L_j  (Y) ] =  \Tr[  (\map L_j^\dag  (X))^\dag \,  Y]\, , \forall X \in  L(\spc K_j),  \forall Y  \in  \spc A$, and, for a subset $ \spc B  \subseteq  \spc A$,   $\Span_\R (\spc B) $ is the linear span of $\spc B$ over the real field.  
  
\end{theo}    
\Proof By Theorem \ref{theo:extreme1},  extremality of $A$ is equivalent to the  injectivity condition  $\forall H  \in \spc V_A  :     \map L_j  (H)  =  0  \, , \forall j\in  \{1,\dots, n\}  \Longrightarrow H =  0$.  Since each map $\map L_j$ is Hermitian-preserving,  $\map L_j( H)$ is a Hermitian operator, and  the condition  $\map L_j  (H)=  0\, ,\forall j$ 
 is equivalent to $\Tr  [  \map L_j  (H)  \,  Y_j]  =  0  \, , \forall Y_j  \in  \Herm(\spc K_j)$, which in turn is equivalent to 
\begin{align}
\Tr  [ H  \,  \map L_j^\dag  (  Y_j)  ]  =  0  \qquad \forall Y_j  \in  \Herm(\spc K_j) \, . 
 \end{align}
 Using the relation   $ H = \Pi_A  H  \Pi_A $, valid for every $H\in\spc V_A$,   we then obtain
 \begin{align}\label{orthogonal}
\Tr  [ \Pi_A  H \Pi_A \,  \map L_j^\dag  (  Y_j)\Pi_\rho  ]=  \Tr  [ H ~\Pi_A   \map L_j^\dag  (  Y_j)\Pi_A ]  =  0  \qquad \forall Y_j  \in  \Herm(\spc K_j) \, .
 \end{align}

Now, the set $\spc V_A$ defined in Eq. (\ref{Arho}) is a vector space over the real field,  and can be regarded as a real Hilbert space with respect to the Hilbert-Schmidt inner product $\<X,  Y  \>  :  =\Tr [  X^\dag Y] $.  Hence,  
$\spc V_A$ can be decomposed as
\begin{align}\label{Drho}
\spc V_A  =  D_A  \oplus  D_A^\perp   \, , 
\end{align} 
 with ${D}_A  :  =    \Span_\R  \Big \{  \Pi_A \map L_j^\dag  (  Y_j) \Pi_A ~\Big|~   Y_j  \in  \Herm  (\spc K_j) \, , j\in  \{1,\dots,  n\} \Big\}^\perp$.  
 The condition $\forall  H\in \spc V_A:  \, \map L_j  (H)   =  0 \, \forall j   \Longrightarrow H = 0$ is then equivalent to  $\forall   H \in \spc V_A  :  \, H \in  D_A^\perp   \Longrightarrow H =0$, that is, $D_A^\perp  = \{0\}$.  By Eq. (\ref{Drho}), this condition is equivalent to   $\spc V_A  =  D_A$. \qed

 In the next section we  will see that the extremality condition  (\ref{extremalitycondition}) sometimes yields   better bounds than Eqs. (\ref{boundC}) and (\ref{boundR}).
  

\section{Applications}\label{sec:applications}  
  
Here we illustrate the application of our general results to  special cases, retrieving known results such as Choi's characterisation of the extreme quantum channels \cite{choi1975completely}, Parathasarathy's characterisation of the extreme bipartite states with given marginals  \cite{parthasarathy2005extremal}, and the quantum version of Birkhoff theorem for qubit bistochastic channels \cite{landau1993birkhoff,mendl2009unital}.  
We also provide a few  new results in the real vector space case. 

\subsection{Extreme quantum states with given expectation values}  

Consider the spectrahedron of normalised quantum states $\rho  \in  \Herm_+  (  \spc H)$ with presecribed expectation values of a set of observables $O_j\, ,  j\in \{1,\dots,  n-1\}$, as defined in Eq. (\ref{givenexpectationvalues}).    In this case,   $\spc V_\rho   =  \Herm  (\Supp  (\rho))$  and the extremality condition (\ref{extremalitycondition}) in Theorem \ref{theo:extreme} reads
\begin{align}\label{extremestates}
  \Herm  (\Supp  (\rho))  =  \Pi_\rho\,  \Big( \Span_\R  \big\{    O_j ~\big|~  j\in\{1,\dots,  n\}  \big\}  \Big)  \,   \Pi_\rho     \, ,  
\end{align} 
where $\Pi_\rho$ is the projector on the support of $\rho$ and $O_n:  =    I$.  

In the standard case where $\spc H$ is a complex Hilbert space, the bound (\ref{boundC}) reads
\begin{align}\label{complexstates}
r^2  \le n  \, ,
\end{align}
where $r$ is the rank of $\rho$.  This bound implies that  the extreme points are rank-one whenever $n\le 3$.   To have rank-two extreme points,  we need  at least $n=4$.    This condition is tight:  for example, the density matrix $I/2  \in  \Herm_+(\C^2)$  is an extreme point (in fact, the only point) of the convex set of normalised density matrices  with zero expectation values of the three Pauli observables  $  X:=  \begin{pmatrix} 0& 1  \\  1&0\end{pmatrix}$,  $  Y:=  \begin{pmatrix} 0& -i  \\  i&0\end{pmatrix}$,   and  $  Z:=  \begin{pmatrix} 1& \phantom{-}0  \\  0&-1\end{pmatrix}$.  More generally, every normalised  density matrix $\rho \in  \Herm_+ (\C^2)$  is an extreme point  of the convex set of normalised density matrices with a given expectation value of the Pauli observables $X, Y,Z$.

In the real vector space case, the bound (\ref{boundR}) reads  
\begin{align}\label{realstates}
\frac{r(r+1)}2    \le  n   \, .
\end{align}  
  This bound implies that the extreme points are rank-one whenever $n\le 2$.   To have rank-two extreme points,  one needs  at least $n=3$.    Also this condition is tight:   for example, the density matrix $  I/2\in \Herm_+(\R)$ is an extreme point (in fact, the only point) of the set of real-valued density matrices  with zero expectation values of the Pauli matrices  $X$ and $Z$.  
 More generally, every  normalised density matrix $\rho  \in  \Herm_+  (\R^2)$ is an extreme point (in fact, the only point) of the convex set of real-valued normalised  density matrices with a given expectation value of the Pauli observables $X$ and $Z$.

\subsection{Extreme phase-covariant POVMs,  Hadamard channels, and correlation matrices}\label{subsec:hadamard} 

Phase-covariant POVMs, introduced  in Holevo's work on statistical decision theory \cite{holevo1973statistical,holevo2011probabilistic},  have found many applications in quantum metrology, quantum communication,   and other areas of quantum information theory.  Consider the unitary operators  $U_\theta   =  \sum_{j=1}^{n}   \,e^{i n_j  \theta} \,  |j\>\<j|  \in  L(\spc H)$, with $\spc H = \C^n$,  $\theta \in  [0,2\pi)$, and $n_j  \in  \Z, \, \forall j\in\{1,\dots, n\}$, $n_j  \not  = n_l \, , \forall j\not  =  l$.    In the simplest instance, a phase-covariant POVM is an operator valued measure of the form $M_\xi(\d \theta)= U_\theta  \xi  U_\theta^\dag\,  \d \theta/(2\pi)$, where $\xi  \in  \Herm_+  (\spc H)$ is a positive operator satisfying the normalisation condition  
\begin{align}
\int_0^{2\pi}  \frac{\d \theta}{2\pi}  ~U_\theta  \xi  U_\theta^\dag    =  I \, ,  \qquad { \rm or~equivalently} \qquad  \<j| \xi|j\>   =  1  \quad \forall j\in\{1,\dots,  n\} \,  . 
\end{align} 
Covariant POVMs form a convex set. Since the covariant POVM $M_\xi(\d \theta)$ is completely described by the operator $\xi$, the   convex set of covariant POVMs is in one-to-one with the spectrahedron 
\begin{align}
{\sf S}_{\rm corr}   :=  \big\{  \xi \in  \Herm_+ (\spc H)~\big|  ~ \<j|\xi|j\> =1 \, \forall j\in  \{0,\dots, d-1\}  \big\} \, .
\end{align}   
This spectrahedron is also known as the spectrahedron of {\em (normalised) correlation matrices} \cite{grone1990extremal,li1994note,kiukas2008note} or as an {\em elliptope} \cite{laurent1995positive}.    
Up to a proportionality factor, this spectrahedron coincides with the spectrahedron of all quantum states  $\rho \in  \Herm_+ (\spc H)$ with fixed expectation values $\Tr  [\rho \, O_j]=1/n$ of the observables $O_j:= |j\>\<j| \, ,  j\in \{1,\dots,  n\}$. 

The spectrahedron of phase-covariant POVMs/correlation matrices is also isomorphic to the set of completely positive trace-preserving maps $ \map C_\xi:  L (\spc H ) \to L(\spc H) $ of the form  
\begin{align}\map C_\xi  (A)    :=   \xi \circ A\,,
\end{align}  where $\xi \circ A$ is the Hadamard product defined by the relation  $\<  j|   \,  (\xi \circ A)  \, |l\> : =  \< j|  \xi  |l\>  \,  \<j|  \xi  |l\>$  \cite{buscemi2005inverting}.    These maps are also known as  {\em quantum Hadamard channels} \cite{king2005properties,bradler2010trade,winter2016potential} in the quantum information literature.    

 Physically,  every phase-covariant POVM can be obtained by  applying a Hadamard channel to the system, and then measuring a fixed phase-covariant POVM: for every quantum state $\rho  \in \Herm_+  (\spc H)$ and every measurable set $B \subseteq [0,2\pi)$, one has
\begin{align}
\Tr  [ M_\xi(  B) \,  \rho ]    =  \Tr  [M_{\xi_0}  (B) \,  \map C_\xi  (\rho)]  
\end{align}  
where $\xi_0$ is the positive operator  with  $\<j|  \xi_0|l\>=1$,  $\forall  j,l \in  \{1,\dots,  n\}$.   

The extreme points of the convex sets of correlation matrices/covariant POVMs/Hadamard channels have been characterised by Li and Tam in Ref. \cite{li1994note}. This characterisation can be obtained from condition (\ref{extremestates}), as shown in the following: 
\begin{prop}\label{prop:extremecor}
Let   $\xi  \in  L(\spc H)$ be a point of the spectrahedron of correlation matrices, with $\spc H=  \K^n$ for either $\K  = \C$ or $\K=\R$,   let $X :     \spc H \to \K^r$ be a linear operator such that $\xi  =  X^\dag X$, and let $|x_j\>:  =  X|j\>   \in \K^r \, , j\in  \{1,\dots,  n\}$ be the columns of $X$.       Then, $\xi$ is  extremal if and only if 
 \begin{align}\label{extremecor}
  \Herm  (\K^r)  =      \Span_\R  \big\{   |x_j\>\<x_j|   ~\big|~  j\in\{1,\dots,  n\}  \big\}        \, ,  
\end{align} 
\end{prop}  

\Proof   The map $\map X :  L(   \Supp (\xi)   ) \to  L(\K^r),  A \mapsto \map X  (A)   :=X  A  X^\dag$ is invertible and Hermitian-preserving.  Hence, the extremality condition (\ref{extremestates}) is equivalent to 
 \begin{align}
  \Herm  (\K^r)  =      \Span_\R  \big\{   X \Pi_\xi \, O_j  \,  \Pi_\xi  \,X^\dag ~\big|~  j\in\{1,\dots,  n\}  \big\}       \, ,  
\end{align} 
where $\Pi_\xi$ is the projector onto the support of $\xi$.  Recalling the relations $  X\Pi_\xi   =  X$ and
   $O_j  =  |j\>\<j|\,, \forall j\in\{1,\dots,  n\}$ we then obtain Eq. (\ref{extremecor}).   \qed  

Extremal correlation matrices can be constructed from Proposition (\ref{prop:extremecor}) by first finding a  set of unit vectors $(|x_j\>)_{j=1}^n  \in  \K^r$ such that the projectors  $(  |x_j\>\<x_j|)_{j=1}^n$ span $\Herm  (\K^r)$, with $\K  =\R$ or $\K  = \C$, and then defining $\xi_{jl}:  =  \<x_j|  x_l\>$.      For example, one can pick the three vectors $\{|x_1\>  :  = |1\>  ,  |x_2\>  := |2\>  , |x_3\>: =  (|1\>+  |2\>)/\sqrt 2\}\subset \R^2$  and the four vectors  $\{|x_1\>  :  = |1\>  ,  |x_2\>  := |2\>  , |x_3\>: =  (|1\>+  |2\>)/\sqrt 2,  |x_4\>:  =  (|1\>  + i|2\>)/\sqrt 2\}\subset \C^2$ and construct the rank-two extreme correlation matrices
 \begin{align} 
 \xi  =  \begin{pmatrix}  1  &   0   &  \frac 1{\sqrt 2}  \\  0  &  1   & \frac 1{\sqrt 2} \\
\frac 1{\sqrt 2}   &\frac 1{\sqrt 2}  & 1 \end{pmatrix}    \in   M_3 (\R)  
 \qquad {\rm and}  \qquad  
 \xi  =  \begin{pmatrix}  1  &   0   &  \frac 1{\sqrt 2}    &  \frac 1{\sqrt 2}  \\  0  &  1   & \frac 1{\sqrt 2}    &  \frac i{\sqrt 2} \\
\frac 1{\sqrt 2}   &\frac 1{\sqrt 2}  & 1   &  \frac{  1+  i}2 \\
  \frac 1{\sqrt 2} &  \frac {-i}{\sqrt 2}  &    \frac {1-i}{ 2}  & 1  
\end{pmatrix}    \in   M_4 (\C)   
\end{align}
More generally, any three pure quantum states with non-collinear Bloch vectors in the real case, or every four pure quantum states with non-coplanar Bloch vectors in the complex case can be used to construct a rank-two extreme correlation matrix in $M_3(\R)$ or in $M_4 (\C)$, respectively. 

Note that the size of the matrix in the above examples is minimal:   bounds (\ref{boundC}) and (\ref{boundR}) imply that 
 the rank $r$  of an extremal correlation matrix satisfies the condition $r^2  \le n$ in the complex case, and $r(r+1)/2 \le n$ in the real case \cite{li1994note}.   In terms of the associated Hadamard channels, we have the following corollary: 
 \begin{cor}[\cite{buscemi2005inverting}]\label{cor:unitaryextr}
The extreme points of the set of Hadamard channels  on a $d$-dimensional quantum system with complex (real) Hilbert space are unitary  channels whenever $d  <4$  ($d<3$).  
 \end{cor}

\subsection{Extreme POVMs and extreme state ensembles}  
   For a quantum system with Hilbert space $\spc H$,  the statistics of a given measurement with outcomes in a finite set  $\set X$  is described by a POVM, corresponding to a family of operators $P_x \in\Herm_+(\spc H) \,,  x\in\set X$ satisfying the normalisation condition  $\sum_{x\in \set X} \, P_x  =  I$.     The POVM  $(P_x)_{x\in\set X}$ in $\Herm_+(\spc H)$ can be equivalently represented as a single operator  $P:  =  \bigoplus_{x\in\set X}   P_x$ in the finite dimensional algebra  $\spc A   :  = \bigoplus_{x\in\set X}  L(\spc H_x)$,  $\spc H_x\simeq \spc H\, , \forall x\in\set X$.  The convex set of all POVMs on $\spc H$ with outcomes in $\set X$ is then equivalent to the spectrahedron  consisting of positive operators $P  \in \spc A_{\Herm,  + }$ satisfying the condition $\map L(P): =  I$ where $   \map L:  \spc A \mapsto  L(\spc H)$ is the Hermitian-preserving map defined by $\map L (\bigoplus_x  P_x) : =\sum_{x\in\set X}  P_x $.   Various characterisations of the extreme points of this spectrahedron have been provided in Refs. \cite{parthasarathy1999extremal,winter2004extrinsic,dariano2005classical,chiribella2007continuous,chiribella2010barycentric,farenick2011classical}.

A spectrahedron with similar structure is the spectrahedron of  the ensemble decompositions of a given quantum state \cite{hughston1993complete}, which plays a central role in the evaluation of the classical capacity of quantum channels \cite{holevo1973bounds,davies1978information,sasaki1999accessible}.  An ensemble decomposition of a density matrix $\rho  \in  \Herm_+ (\spc H)$ with alphabet $\set X$ is a collection of operators $ \rho_x \in \Herm_+  (\spc H) \, , x\in\set X$ such that $\sum_{x\in\set X}  \rho_x  =  \rho$. The set of ensemble decompositions of the state  $\rho$ with  a given alphabet $\set X$  is convex and is in one-to-one correspondence with the spectrahedron consisting of all positive operators $P \in \spc A_{\Herm, +}$ satisfying the condition $\map L(P)  = \rho $, where $  \spc A$ and $\map L$ are defined as in the previous paragraph.

In general, we define the spectrahedron  
\begin{align}{\sf S}_\Gamma:=  \bigg\{  \bigoplus_{x\in\set X}  P_x    \in\spc A_{\Herm ,  +} ~\bigg|~ \sum_{x\in\set X}  P_x   =  \Gamma \bigg\} \,,
\end{align}  
for some positive operator $\Gamma$.  In this way, the spectrahedron of POVMs corresponds to $\Gamma  =  I$, whereas a spectrahedron of ensemble decompositions corresponds to $\Gamma=  \rho$, for some  normalised density matrix $\rho$.
 
 Theorem \ref{theo:extreme1}  implies the following characterisation: 
 \begin{prop}\label{prop:povm1}
 Let $P  =  \bigoplus_{x\in\set X}   P_x$ be an element of ${\sf S}_\Gamma$, and, for every $x\in\set X$, let $\{  H^{(x)}_m\}_{m=1}^{r_x}$ be a basis for $\Herm  (\Supp(P_x))$.  Then,  $P$ is extremal if and only if the operators $\left(  H^{(x)}_{m}\right)_{x\in\set X  ,  m\in  \{1,\dots,  r_x\} }$  are linearly independent.
  \end{prop}
 \Proof  The extremality condition in Theorem (\ref{theo:extreme1}) is that, for every family of operators $H_x\in \Herm  (\Supp(P_x))\, ,  x\in\set X$, the condition $\sum_x  H_x  = 0 $ implies $H_x=  0\, \forall x\in\set X$.   Now, an operator  $H_x$ is in $\Herm  (\Supp(P_x))$ if and only if it can be decomposed as $H_x=  \sum_{m}  \,  c^{(x)}_m     \,   H_m^{(x)} $ for  real coefficients $\{ c_m^{(x)}\}_m \subset \R$.   Hence, the extremality condition becomes that, for every tuple of real coefficients $(c_m^{(x)})_{x,m}$ the condition, $\sum_x\sum_{m}  \,  c^{(x)}_m     \,   H_m^{(x)} =0 $ implies $c_m^{(x)} =  0$ for every $x$ and $m$. \qed      
 
 In particular, Proposition \ref{prop:povm1}  implies the following corollary, proven in \cite{dariano2005classical} for POVMs in the complex case: 
\begin{cor}
Let $P  =  \bigoplus_{x\in\set X}   P_x$ be an element of ${\sf S}_\Gamma$, and, for every $x\in\set X$, let $\{ |\psi_m^{(x)}\> \}_{a=1}^{r_x}$ be a  basis for  $\Supp (P_x)$. 
In the complex case, $P$ is extremal if and only if the operators $\left( O_{ab}^{(x)}   :=  |\psi_a^{(x)}\>\<  \psi_b^{(x)}| \right)_{x\in\set X \, , 1\le a \le r_x,  ~  1\le  b\le r_x}$
are linearly independent over $\C$.   In the real case,  $P$ is extremal if and only if the operators $   \left({\rm Sym}  (   O_{ab}^{(x)} :  =        (O_{ab}^{(x)}   +  O_{ba}^{(x)})/2  ) \right)_{  x\in\set X \, , 1\le a \le b\le r_x}$
are linearly independent over $\R$.   
\end{cor} 
\Proof   In the real case,  it is enough to  apply Proposition \ref{prop:povm} to  the operators  $\left\{{\rm Sym}  (O_{ab}^{(x)})\right\}_{ 1\le a \le b  \le r_x}$, which form a basis for $\Herm (\Supp  (P_x))$.   In the complex case, a  basis of   $\Herm (\Supp  (P_x))$  is provided by the operators $\left\{{\rm Sym}  (O_{ab}^{(x)})\right\}_{ 1\le a \le b  \le r_x} \cup  \left\{{\rm Ant}  (O_{ab}^{(x)}):   =   (O_{ab}^{(x)}   -  O_{ba}^{(x)})/(2i)    \right\}_{ 1\le a   < b  \le r_x}$.   Their linear independence over $\R$ is equivalent to the linear independence of the operators $\left(   O_{ab}^{(x)}  \right)_{x\in  \set X,  1\le a\le r_x \, , 1\le b\le r_x}$ over $\C$. \qed   

 An alternative characterisation can be obtained from Theorem \ref{theo:extreme}:     an element $P  \in  {\sf S}_\Gamma$  is extremal if and only if  
\begin{align}\label{povmspan}
\Herm  (\Supp (P))   =  \Span_\R  \Big\{   \bigoplus_{x\in\set X}    \Pi_{x}  \, A \,  \Pi_{x}    \Big|  \,   A \in \Herm(\spc H)  \Big\}   \, ,
\end{align}
where $\Pi_x$ is either the projector on  $\Supp (P_x)$. For POVMs, this condition was proven in \cite{chiribella2010barycentric}.

Proposition \ref{prop:povm1} and Eq. (\ref{povmspan}) imply that  the ranks of  operators $(P_x)_{x\in \set X}$ satisfy the bounds  
\begin{align}
\sum_{x\in\set X}  \frac{ r_x (r_x+1)}2    \le  \frac {d(d+1)}2    \qquad {\rm and}  \qquad   \sum_{x\in\set X}  r_x^2    \le   d^2 \, ,
\end{align}
in the real and complex case, respectively. These two bounds imply that the number of nonzero operators is at most $f_\K (d)$, with $f_\R (d)=d(d+1)/2$  and $f_\C  (d)  =  d^2$.      A consequence of this fact is that  the optimal POVM (ensemble)  for any convex optimisation problem  can be taken without loss of generality to have at most $f_\K  (d)$ outcomes  (states).   


 \subsection{Extreme quantum states with fixed marginals}  

Consider the spectrahedron   (\ref{givenmarginals}) consisting of all $k$-partite quantum states $\rho  \in  \Herm_+  (\spc H_1\otimes \cdots \otimes \spc H_k)$ with marginals  $(\rho_{S_1},  \rho_{S_2},  \dots,  \rho_{S_n})$  on subsystems $(S_1,\dots,  S_n)$, with $S_j  \subseteq \{  1,\dots,  k\} \, \forall j\in \{  1,\dots,  n\}$.   By Theorem \ref{theo:extreme},  an element $\rho$ of this spectrahedron is extremal if and only if
\begin{align}\label{extrememarg}
\Herm  (\Supp  (\rho))    =   \,  \Pi_\rho  \left[  \Span_
\R \Big(   \big(   \Herm  (\spc H_{S_j})\big)  \otimes I_{\overline S_j}  \Big)_{j=1}^n \right]   \Pi_\rho \, ,
\end{align}
 where $\Pi_\rho$ is the projector on the support of $\rho$ and $\spc H_{S_j}   =  \bigotimes_{k\in S_j}  \spc H_k$. In the complex case, Eq. (\ref{extrememarg}) implies that  the rank of an extreme point $\rho$ can be upper bounded as 
 \begin{align}\label{inclusionexclusion}
 r^2  \le  \sum_{j=1}^n    \,    d_{S_j}^2   -   \sum_{1\le j  <  k\le n}   \,  d^2_{S_j\cup S_k}  +  \sum_{1\le j  <  k<l\le n}   d^2_{S_k\cup S_k\cup S_l}       +  \dots  +  (-1)^{n+1}  \,  d^2_{S_1\cup S_2\cup \cdots \cup S_n} \, ,
 \end{align}
   where  we used the notation $ d_{\emptyset}: = 1$ and  $d_S :  =\prod_{k \in  S}  d_k$, for an arbitrary non-empty  subset $S  \subseteq \{1,\dots,  n\}$.  Eq. (\ref{inclusionexclusion}) follows by applying the inclusion-exclusion principle to the bases    $B_{j}  :  =  \left\{  \bigotimes_{l \in  S_j}  \bigotimes_{m_l=1}^{d_l^2}   H_{m_l}^{(l)}  \otimes   I_{\overline S_j}   \right\}$,  $j\in \{1,\dots,  n\}$, where, for every  $l$,  $\big\{  H^{(l)}_{m_l} ~\big|~ m_j  \in  \{1,\dots,  d_l^2\}\big\}$ is a basis for $\Herm  (\spc H_l)$.  
   

 Now, consider the special case of the spectrahedron  consisting of all bipartite states on  $\spc H_1\otimes \spc H_2$ with   marginals $\rho_1$ and $\rho_2$  on the Hilbert spaces $\spc H_1$ and $\spc H_2$, respectively.  This spectrahedron, hereafter denoted by ${\sf S}  (  \rho_1,  \rho_2)$  was studied by  Parthasarathy  \cite{parthasarathy2005extremal} and Rudolph  \cite{rudolph2004extremal}.   Parthasarathy's characterisation coincides  with Eq. (\ref{extrememarg}) in the case $n=2$,  $S_1  =  \{1\}$ and $S_2  =  \{2\}$.     Rudolph's characterisation coincides with the Theorem \ref{theo:extreme1},  applied to the spectrahedron ${\sf S}(\rho_1,\rho_2)$, and will be discussed in subsection \ref{subsec:channels}.     Both authors  also gave the  bound 
   \begin{align}\label{cmarg}
  r^2   \le  d_1^2  +  d_2^2  -1 \, ,
  \end{align}
which  coincides with  Eq. (\ref{inclusionexclusion}) in the special  case $n=2$,  $S_1  =  \{1\}$ and $S_2  =  \{2\}$.  In this case, Eq. (\ref{extrememarg}) also implies a bound in the real case,  namely 
 \begin{align}\label{rmarg}
  \frac{r (r+1)}2   \le  \frac{d_1 (d_1+1)}2  +  \frac{d_2 (d_2+1)}2  -1 \, ,
  \end{align}
  For $d_1  =  d_2  = 2$,  Eqs. (\ref{cmarg}) and (\ref{rmarg})  imply that  the rank of the bipartite state $\rho$ cannot be larger than two.    
    If in addition $\rho_1  =  \rho_2=  I/2$,  then   all extreme states  have rank one, and therefore correspond to maximally entangled states:
    \begin{prop}[\cite{parthasarathy2005extremal,rudolph2004extremal}]\label{prop:extrememaxent}
A bipartite state $\rho$ is an extreme point of ${\sf S}  \left(\frac I2,  \frac I2\right)$ if and only if $\rho$  is rank-one, that is, if and only if it is a maximally entangled state. 
 \end{prop}  
 For convenience of the reader, we provide a proof  in Appendix \ref{app:maxent}.    We also observe that, via the Choi isomorphism,  the above proposition is equivalent to the statement that every bistochastic qubit channel is a  convex combination of unitary channels,  a fact known as the quantum Birkhoff theorem \cite{landau1993birkhoff}.  
   
One may wonder whether an analogue of Proposition \ref{prop:extrememaxent}   holds for the extreme points of every set ${\sf S}  (\rho_1,\rho_2)$, consisting of bipartite quantum states with equal marginals $\rho_1   = \rho_2$.   The answer is negative: as a counterexample, take $\rho_1 =  \rho_2  =  p\,  |0\>\<0| +  (1-p)  \,  |1\>\<1|$ with $p\in  (0,1/2)$. All the rank-one states  $ \rho \in  {\sf S}  (\rho_1,\rho_2)$ are of the form   $\rho  =  |\Psi\>\,\Psi|$ with $|\Psi\>  =  c_0  \,  |0\>\otimes |0\>  +  c_1  \,  |1\>\otimes |1\>$, with $|c_0|^2  =  p$ and $|c_2|^2  =  1-p$.  Note that all these states satisfy the condition $(\<0|  \otimes \<1|) \,  \rho  \,  (|1\> \otimes |0\>)=  0$  Now,     consider the rank-two state  
\begin{align}
\rho_*   =  \frac {  |\psi_+\>\<\psi_+  |  \otimes |\psi_+\>\<\psi_+  |   +   |\psi_-\>\<\psi_-  |  \otimes |\psi_-\>\<\psi_-  |}2\,,  
\end{align}  
where  $|\psi_{\pm}\>  := \sqrt{ p }   |0\>  \pm  \sqrt{1-p} \, |1\>$.  
Since $(\<0|  \otimes \<1|) \,  \rho_*  \,  (|1\> \otimes |0\>)  =  p(  1-p) \not  =  0$, $\rho_*$ cannot be a convex combination of rank-one extremal points.

\subsection{Extreme completely positive maps subject to  Hermitian mapping constraints}\label{subsec:channels} 

As we have seen in Section \ref{sec:framework}, the sets of quantum channels, bistochastic/Gibbs-preserving/energy-preserving channels, and other sets of quantum maps are examples of convex sets  of completely positive maps subject to Hermitian mapping constraints.  To treat these sets in a unified way, here we consider the spectrahedron  $\sf S_{\rm map}$  defined in Eq.  (\ref{mappingconstraints}), with Hermitian operators $\{A_j\}_{j=1}^n  \subset \Herm  ( \spc H_{\rm in})$ and $\{  B_j\}_{j=1}^n  \subset \Herm  (\spc H_{\rm out})$.

 Theorem \ref{theo:extreme1} provides a general characterisation of the extreme points, both in the complex and in the real case:  
 \begin{prop}\label{prop:extrememapping}
 Let $C=  \sum_{a=1}^r   |\Gamma_a\>\<\Gamma_a|$ be an element of ${\sf S}_{\rm map}$, with linearly independent vectors $\{|\Gamma_a\>\}_{a=1}^r$, and, for every $a\in  \{1,\dots,  r\}$, let $C_a:  \spc H_{\rm in}\to \spc H_{\rm out}$ be the operator such that $|\Gamma_a\>    =    (I_{\rm in}\otimes  C_a) \,  |\Phi\> \,,$ with $|\Phi\>:  =  \sum_{l=1}^{d_{\rm in}} \,  |l\>\otimes |l\>$.    For complex Hilbert spaces $\spc H_{\rm in}$ and $\spc H_{\rm out}$, the operator $  C$   is an extreme point of ${\sf S}_{\rm map}$ if and only if the operators  
 \begin{align}\label{extrememapping}
\left(   
\left[ \bigoplus_{j=1}^k  C_a    A_j   C_b^\dag    \right]
\oplus
\left[  \bigoplus_{j=k+1}^n  C_a^\dag   B_j\,   C_b\right]  
\right) _{1\le a \le r, \,   1\le b\le r}     
\end{align} 
are linearly independent over $\C$.   

   For real Hilbert spaces $\spc H_{\rm in}$ and $\spc H_{\rm out}$,  let $ {\rm Sym}  (A) :  =  A+  A^T/2$ denote the symmetric part of an arbitrary operator $A$ in    $L(\spc H_{\rm})$ or in $\spc H_{\rm out}$. Then,   the operator $  C$   is an extreme point of ${\sf S}_{\rm map}$ if and only if the operators  
 \begin{align}
\left(   
\left[ \bigoplus_{j=1}^k {\rm Sym} (C_a    A_j   C_b^T)    \right]
\oplus
\left[ \bigoplus_{j=k+1}^n  {\rm Sym}(C_a^T   B_j\,   C_b)\right]  
\right) _{1\le a\le b\le r}       
\end{align} 
are linearly independent over $\R$.   
 \end{prop}
 The proof is provided in Appendix \ref{app:extrememapping}.
   
Proposition \ref{prop:extrememapping}  implies several old and new results.   In the complex case, it implies:
\begin{itemize}
\item {\em Choi's characterisation of the extremal quantum channels \cite{choi1975completely}.}   It corresponds to the case $k=0$, $n=1$,  $A_j=  I_{\rm in}$,  $B_j  =  I_{\rm out}$.     In this case,  Eq. (\ref{extrememapping}) reduces to the requirement that   the operators  $  (C_a^\dag  C_b)_{1\le a\le r,  1\le b\le r}$ be linearly independent.  
\item {\em Landau-Streater's characterisation of the extremal bistochastic channels \cite{landau1993birkhoff}.}   It corresponds to the case $k=1$, $n=2$,  $A_1= A_2  = I_{\rm in}$, $B_1  =  B_2    =  I_{\rm out}$.     In this case,  Eq. (\ref{extrememapping}) reduces to the requirement that   the operators  $  (C_a  C_b^\dag  \oplus  C_a^\dag C_b)_{1\le a\le r,  1\le b\le r}$ be linearly independent.   Moreover, Proposition \ref{prop:extrememaxent} implies that all the extremal bistochastic channels on two-dimensional quantum systems are unitary, a fact known as the quantum Birkohoff theorem  \cite{landau1993birkhoff,mendl2009unital}. 
\item {\em A characterisation of the extremal Gibbs-preserving  channels.}  It corresponds to the case $k=1, n=2, \, A_1= \rho_{\rm in}, \,   A_2  = I_{\rm in},  \, B_1  =\rho_{\rm out}$,    $B_2    =  I_{\rm out}$.     In this case,  Eq. (\ref{extrememapping}) reduces to the requirement that   the operators  $  (C_a   \rho_{\rm in} C_b^\dag   \oplus  C_a^\dag C_b)_{1\le a\le r,  1\le b\le r}$ be linearly independent. 
\item {\em A characterisation of the extremal channels that transform a given input observable into a given output observable.}     It corresponds to the case in which  $k=0$ and $A_j= P_j$ and $B_j  =  Q_j$, for two POVMs $(P_j)_{j=1}^n$ and $(Q_j  )_{j=1}^n$.      In this case,  Eq. (\ref{extrememapping}) reduces to the requirement that   the operators  $  \left( \bigoplus_{j=1}^n C_a^\dag Q_j  C_b \right)_{1\le a\le r,  1\le b\le r}$ be linearly independent.  

\item {\em Rudolph's characterisation of the extremal bipartite quantum states with given marginals on the two systems   \cite{rudolph2004extremal}.}  This case corresponds to  $k=1$, $n=2$,  $A_1   =I_{\rm in}/d$  $A_2  =  \rho_1$,  $B_1  = \rho_2  $, $B_2  =  I_{\rm out}/d$.    Like in the case of bistochastic channels,  the requirement for extremality is that  the operators  $  (C_a  C_b^\dag  \oplus  C_a^\dag C_b)_{1\le a\le r,  1\le b\le r}$ be linearly independent.
\end{itemize} 
 Similar results can be obtained in the real case, replacing  each of the operators in the above sets with its symmetrised version, and restricting the range of the indices $a$ and $b$  to $1\le a \le b\le r$.

\subsection{Extreme quantum combs}
  A quantum process receiving inputs and outputs at $n$ time subsequent steps is described by a quantum channel  $\map C:   \Lin  \left( \bigotimes_{m=1}^n \spc H^{(m)}_{\rm in}\right)   \to   \Lin  \left( \bigotimes_{m=1}^n \spc H^{(m)}_{\rm out}\right)$   with the property that the marginal state of the first $j$ output systems does not depend on the input systems state of the remaining $n-j$ systems.  Multitime processes of this form are also known as quantum channels with memory \cite{kretschmann2005quantum},  quantum strategies \cite{gutoski2007toward}, and quantum combs \cite{chiribella2008quantum,chiribella2009theoretical}.

 In the Choi representation, quantum combs correspond to the spectrahedron  
 \begin{align}
 {\sf S}_{\rm comb}     :  =  & \bigg\{  C  \in  \Herm_+ \left(\bigotimes_{m=1}^n \spc H_{\rm in}^{(m)} \otimes \spc H_{\rm out}^{(m)}\right)  ~\bigg|  ~ \Tr_{ \rm out   }  [C]   =  I_{\rm in}      \label{channelagain}   \\
  & \quad \Tr_{{\rm out }   > j }     [C ]    =    \frac{ I_{{\rm in}  >  j} }{d_{{\rm in}  >  j}} \otimes     \Tr_{ {\rm in}  >j  \,  , {\rm out }   > j   }[      C ]   \, ,   \quad \forall j\in  \{1,\dots,  n-1\}\bigg\}
\label{spectracomb}  
\, ,
 \end{align}
where  $\Tr_{{\rm out}  >j}$   ($I_{{\rm in}  >j} $) is the partial trace  (identity operator) over  $\spc H_{{\rm out}  >j} :  =\bigotimes_{m>j}  \spc H_{\rm out}^{(m)}$   ($\spc H_{{\rm in}  >j} :  =\bigotimes_{m>j}  \spc H_{\rm in}^{(m)}$),    $\Tr_{\rm out} : =\Tr_{{\rm out}  >0}$  ($I_{\rm in}   :  =I_{{\rm in}\,  > 0}$) is the partial trace (identity operator) over $\spc H_{\rm out}:  =  \spc H_{{\rm out}  >  0}$  ($\spc H_{\rm in}:  =  \spc H_{{\rm in}  >  0}$),  $\Tr_{{\rm in}  > j ,  {\rm out}  >j}$ is the partial trace over $\spc H_{{\rm in}  >j} \otimes \spc H_{{\rm out}  >j}$ and $d_{{\rm in }  >  j}  :=  \prod_{m>j}   d_j$.  
In Eq. (\ref{spectracomb}) it is understood  that the order of the tensor factors in the right-hand-side of the above equation is suitably rearranged to match the order of the tensor factors on the left-hand-side.    

 Conditions for extremality in   $ {\sf S}_{\rm comb} $  have been provided  in Refs.  \cite{dariano2011extremal,jenvcova2013extremal}.    We now provide an equivalent characterisation,  obtained from Theorem \ref{theo:extreme}.  

\begin{prop}
An operator $C  \in  \Herm_+ \left(\spc H_{\rm in} \otimes \spc H_{\rm out}\right)$ in the spectrahedron of quantum combs is extremal if and only if 
\begin{align}
\Herm  (\Supp  (C))   =  &  \Pi_C  \Span_\R  \Bigg  \{   I_{\rm out}  \otimes \Herm  (\spc H_{\rm in} )  \\
       & \cup  \Bigg(\bigcup_{j=1}^{n-1}      I_{{\rm out} \,   >j}    \otimes     \Herm  (  \spc H_{{\rm in}  \le j}  \otimes \spc H_{{\rm out}  \le j}) \otimes  \Herm_0  (  \spc H_{{\rm in}  >  j})   \Bigg)    \Bigg \}    \Pi_C  \, , 
\end{align}
where $\Pi_C$ is the projector onto the support of $C$ and  $\Herm_0 (\spc H)  \subset  \Herm(\spc H)$ is the space of traceless Hermitian operators on a given Hilbert space $\spc H$. 
\end{prop}

\Proof Theorem \ref{theo:extreme} implies that an element $C$ of the spectrahedron of quantum combs is extremal if and only if 
\begin{align}\label{comb1}
\Herm  (\Supp  (C))   =  \Pi_C \,  \Span_\R  \left(   \bigcup_{j=0}^{n-1}   S_j  \right)\,  \Pi_C  \,, 
\end{align} 
where $\Pi_C$ is the projector on $\Supp (C)$,  
\begin{align}
\nonumber S_0 &:=  \big\{   I_{\rm out}  \otimes H~|~  H \in\Herm  (\spc H_{\rm in}) \big\}  \\ 
\nonumber S_j&: =   \Span_\R  \bigg\{ I_{{\rm out} \,   >j}  \otimes   \left(  H  -  \frac{   I_{{\rm in} \,   >j}  \otimes \Tr_{{\rm in} \,   >j}    [H]}{d_{{\rm in}  >j}}\right)   ~\bigg|~   H \in   \Herm (    \spc H_{ {\rm in} }   \otimes \spc H_{{\rm out}   \le j}) \bigg\}  \\
 &\qquad  \qquad  \qquad  \qquad   \qquad  \qquad  \qquad  \qquad  \qquad  \qquad  \qquad      \qquad  \qquad    \forall  1 \le  j <  n\, ,
\end{align} 
with $ \spc H_{{\rm out}   \le j}  :  =  \bigotimes_{m\le j}  \spc H_{\rm out}^{(m)}$.   For $1\le j<n$, the set $S_j$ can be equivalently rewritten as 
\begin{align}
\nonumber S_j   &=  \Span_\R  
\big\{ 
I_{{\rm out} \,   >j}  
\otimes   
T   ~\big|  T   \in   \Herm (    \spc H_{ {\rm in} }  \otimes \spc H_{{\rm out}   \le j} ) \, ,  \Tr_{{\rm in    >j}} [T ]  =  0 \big\}    \\
&  = I_{{\rm out} \,   >j}    \otimes     \Herm  (  \spc H_{{\rm in}  \le j}  \otimes \spc H_{{\rm out}  \le j}) \otimes  \Herm_0  (  \spc H_{{\rm in}  >  j}) \, .
\end{align}
\qed

\section{Spectrahedron-normalised POVMs}\label{sec:POVM}

In this final section we discuss a generalisation of the notion of ``POVM'' commonly used in quantum theory.  In this generalisation, we consider positive operator valued measures that are resolutions of elements of a given spectrahedron.   The main result of the section is the generalisation of a result in Refs. \cite{chiribella2007continuous,chiribella2010barycentric}, stating that every extreme POVM in finite dimensions must have support contained into a finite set of points.

Let $\Omega$ be a measurable space equipped with a sigma-algebra  $\sigma  (\Omega)$ of subsets.  

\begin{defi}[OVMs]
An  {\em operator valued measure  (OVM)} over $\Omega$, with values in an operator algebra $\spc A$,    is a function  $P  :  \sigma  (\Omega)  \to \spc A$ satisfying the countable additivity condition  $P(    \bigcup_{i=1}^{\infty}  \,   B_i )  =  \sum_{i=1}^{\infty}  \,  P(B_i)$ for every sequence $  (B_i)_{i=1}^{\infty} $ with $B_i\in\sigma (\Omega)$ and   $B_i\cap  B_j  =\emptyset  \, ,\forall i\not  =  j$.   
  \end{defi}

\begin{defi}[POVM]
Let $P:  \sigma (\Omega) \to \spc A $ be an  OVM.  If $P(B)$ is a positive operator for every measurable subset $B\in \sigma (\Omega)$, then we say that $P$ is a {\em positive operator valued measure (POVM)}.
\end{defi}

In quantum information, the term POVM  is typically used for measures that are normalised to the identity operator,  namely  
\begin{align}
P(\Omega)   =    I  \,  .
\end{align}
However, other normalisation conditions are sometimes relevant. For example, an ensemble decomposition of a quantum state $\rho \in  \Herm_+  (\spc H)$ can  be described by a POVM $P:   \sigma  (\Omega)  \to  L(\spc H)$ satisfying the normalisation condition 
\begin{align}
P(\Omega)  =   \rho \, .
\end{align} 
A quantum instrument with finite-dimensional input Hilbert space $\spc H_{\rm in}$ and with output Hilbert space $\spc H_{\rm out}$ corresponds via the Choi representation to by a positive operator valued measure $P  : \sigma (\Omega) \to L(\spc H_{\rm in} \otimes \spc H_{\rm out})$   \cite{chiribella2009realization} with the property that $P(\Omega)$ belongs to the spectrahedron of quantum channels, namely  
\begin{align}\label{instru}
\Tr_{\rm out}  [  P(  \Omega)]   =  I_{\rm in} \, .
\end{align} 

Measurements performed through multitime processes  \cite{gutoski2007toward,chiribella2008quantum,chiribella2008memory,ziman2008process,chiribella2009theoretical}  are another example of positive operator valued measures with a different normalisation condition.  In this case, the normalisation condition is
\begin{align} 
P(\Omega)  \in {\sf S}_{\rm comb} 
\end{align}
where $S_{\rm comb}$ is the spectrahedron of quantum combs defined in Eq. (\ref{spectracomb}).  Similar normalisation conditions arise for quantum measurement processes that connect a set of probed processes in an indefinite order\cite{chiribella2019quantum,bavaresco2021strict}.
    
The above examples motivate the following general definition:  
\begin{defi}[Spectrahedron-normalised POVM]
Let  $\spc A$ be an operator algebra, and let $\sf S   \subset  \spc A_{\Herm  ,  +  }$ be a spectrahedron in $\spc A$.    We say that POVM $P:  \sigma (\Omega) \to \spc A $ is {\em ${\sf S}$-normalised} iff $P(  \Omega) $ belongs to $\sf S$.   The set of $\sf S$-normalised POVMs  with outcomes in $\Omega$ will be denoted as  ${\sf POVM}   ( \spc A, \Omega ,  {\sf S})$.
\end{defi}    

We now characterise the extreme points of  the $\sf S$-normalised POVMs with outcomes in $\Omega$ in the case where the algebra $\spc A$ is finite-dimensional and the outcome space $\Omega$ is compact and Hausdorff.     
\begin{theo}[Extreme POVMs have finite support]\label{theo:actuallydiscrete}  
Let $\spc A$ be a finite dimensional algebra,  $D$ be the dimension of the subspace of Hermitian operators in $\spc A$, $\Omega$ be a compact Hausdorff space, and $\sigma (\Omega)$ be the Borel $\sigma$-algebra on $\Omega$.      An element  $P  \in  {\sf POVM}   ( \spc A, \Omega ,  {\sf S})$ is extremal  if and only if  
there exists a finite set $\set X$  with $|\set X|  \le  D$,  an extremal POVM $Q  \in   {\sf POVM}   ( \spc A, \set X,  {\sf S})$, and a function $f:   \set X \to \Omega$   such that
\begin{align}\label{discrete}
P(B)   =   Q(  f^{-1}  (B)  )  \qquad \forall B \in \sigma  (\Omega) \, .
\end{align}   
\end{theo}

\Proof Let $\Supp (P)  :  =\{\omega\in\Omega~|~    P(B)  \not  =   0   \, , \forall B \subseteq \Omega:     B  ~{\rm is~open~and~}  \omega \in  B\}$ be the support of $P$.    To prove the theorem, it is enough to show that $s:=|\Supp (P)|  \le D$.  Indeed, if this condition is satisfied, then one can label the points in $\Supp (P)$ as $(\omega_i)_{i=1}^s$ and, thanks to the Hausdorff property of $\Omega$,   find $s$ disjoint open sets $  B_i  \in \sigma  (\Omega), \,   i\in\{1 , \dots, s\}$ such that $\omega_i  \in  B_i$ for every $i\in  \{1,\dots,  s\}$, and
\begin{align}
P(B)   =  \sum_{i :~ \omega_i \in  B}    \,     P(B_i)    =  Q(  f^{-1}  (B)  ) \, ,
\end{align}     
with $Q(  C ):  =  \sum_{i \in  C}  P_{B_i}\, , \forall  C \subseteq  \{1,\dots,  s\}$, and $ f:  \{1,\dots, s\} \to  \Omega \, ,  i\mapsto  \omega_i$.  

To prove the relation $|\Supp  (P)|\le D$, we proceed by contradiction, following a line of argument used in  Refs.  \cite{chiribella2007continuous,chiribella2010barycentric}.   We observe that the POVM $  P$ satisfies the relation  $P(B)  \le  \mu  (B)  \,  I  \, ,\forall B\in\sigma (\Omega)$, where $\mu  :   \sigma  (\Omega)   \to  \R_+$ is the scalar measure defined by $\mu (B):  =  \Tr  [  P(B)]$.  By Proposition 4.2.2 of \cite{holevo2011probabilistic},  $P$ admits a density with respect to $\mu$, that is,
\begin{align}
P(  B)   =  \int_{B}  \,  \mu  (\d \omega) \,   M(\omega) \,,  \qquad \forall  B\in\sigma (\Omega) \, ,
\end{align} 
where  $M: \Omega \to \spc A_{\Herm ,  +   }$ is a measurable function.  
The function $M$ can be equivalently written as  
\begin{align}
M(\omega)   =   \sum_{j=1}^D   \,   f_j  (   \omega)  \,  H_j  \qquad  f_j(\omega):  =   \Tr  [  H_j \,  M(\omega)] \, ,
\end{align}   
where $\{   H_j\}_{j=1}^D$ be an orthonormal basis for $\spc A_{\Herm    }$ with respect to the Hilbert-Schmidt product.   
Now, let  us assume that $|\Supp (P)|    =  s  >  D$.    Since $\Omega$ is a compact Hausdorff space, the vector space of all measurable functions on $\Omega$ has dimension at least $r$  (see e.g. the proof of Lemma 2 of \cite{chiribella2010barycentric}).   Hence, there exists a measurable function $g  :  \Omega \to  \R$ such that $\<g, f_j\>   :  =  \int  \mu (\d \omega)  \,  g(\omega) \,  f_j(\omega)  =  0 \, , \forall j\in  \{1,\dots,  D\}$.      Without loss of generality, the function $g$ can be chosen to have bounded norm, say $\| g\|   :  = \sup_{\omega \in  \Omega}  |g(\omega)|   \le 1$.   With this choice,  the relation 
\begin{align}
P_\pm  (  B):  =  \int_B  \mu (\d \omega)\,   \big(  1 \pm   g(x)  \big)  \,  M(\omega)  
\end{align}  
defines two POVMs satisfying the condition $P_+   (\Omega)  =  P_{-}  (\Omega)  =   P(\Omega)$.   Hence, both POVMs  $P_\pm$ belong to the convex set ${\sf POVM}   (  \spc A,  \Omega,  {\sf S})$.     Since $P =  (P_+  +  P_-)/2$,  $P$ cannot be extreme, in contradiction with our initial assumption. \qed  

Theorem \ref{theo:actuallydiscrete} reduces the characterisation of the extremal $\sf S$-normalised POVMs with a general compact Hausdorff outcome space $\Omega$ to the characterisation of the extremal  $\sf S$-normalised POVMs with a finite  outcome space $\set X$.   In the finite outcome case, the $\sf S$-normalised POVMs are equivalently represented as finite families  $(P_x)_{x\in \set X}$ of operators $  P_x  \in  \spc A_{  \Herm + }$ satisfying the normalisation condition  
\begin{align}
\sum_{x\in\set X}    P_x   \in  \sf S \,.
\end{align}  
In turn,  ever such family $(P_x)_{x\in \set X}$ can be represented by a single operator $P  : =  \bigoplus_{x\in\set X}  \,  P_x  \in  \widetilde{\spc A}  :  =  \bigoplus_{x\in\set X}  \, \spc A_x $, with $\spc A_x\simeq  \spc A$ for every $x\in\set X$.    
  Hence, the convex set of $\sf S$-normalised POVMs is in one-to-one correspondence with the spectrahedron 
   \begin{align}
  \nonumber    \widetilde{\sf S}    &:  = \Big\{  \bigoplus_{x\in \set X}   P_x     \in    \widetilde{\spc A}_{\Herm,  +}  ~\Big|~  \sum_{x\in\set X}  \,  P_x   \in  {\sf S}  \Big\}  \\
    &   = \Big\{  \bigoplus_{x\in \set X}   P_x     \in    \widetilde{\spc A}_{\Herm,  +}  ~\Big|~    \map L_j    \left(\sum_{x\in\set X}  \,  P_x\right)    =   B_j\, ,   j\in \{1,\dots,  n\} \Big\}   \, ,  \label{sprime}
  \end{align} 
   where $\map L_i  $ ($B_j$) are the maps (operators) defining the spectrahedron $\sf S$.

We conclude the paper with the  characterisation of the extreme points of $\widetilde{\sf S}$.  
\begin{prop}\label{prop:povm}
Let $\spc A$ be a finite dimensional algebra,  $\sf S$ be a spectrahedron in  $\spc A$,  $\set X$ be a finite set, and $\widetilde{\sf S}$ be the spectrahedron defined in Eq.  (\ref{sprime}).  An element $P\in  \widetilde{ S}$ is extremal if and only if the implication  
\begin{align}
\map L_j  \left(   \sum_{x\in\set X}\,    H_x\right)    =  0  \qquad {\Longrightarrow}  \qquad  H_x=  0  ,   \,  \forall x\in\set X    \,     
\end{align}
holds for every set of Hermitian operators $H_x   \in \Herm  (\Supp (P_x))$. 
\end{prop}

The proof is an immediate consequence of Theorem \ref{theo:extreme1}.

The above proposition implies a simple requirement on the operators $(P_x)_{x\in\set X}$:  
\begin{cor}
Let $P  = (P_x)_{x\in\set X}$ be an $\sf S$-normalised POVM with values in a finite dimensional algebra $\spc A$, and, for every $x \in \set X$, let
$\big( |\psi_a^{(x)}\>\big)_{1\le a \le r_x}$ be a basis  of $\Supp (P_x)$.  If $P$ is extremal  and $\spc A$ is an algebra over $\C$,  then the operators $\Big(O_{ab}  :  =  |\psi_a^{(x)}\>\< \psi_b^{(x)} | ~\Big|~  1\le a\le r,  \,  1\le b \le r\, , x\in\set X\Big)$ are linearly independent.
If $P$ is extremal and $\spc A$ is an algebra over $\R$,  then  the operators $\Big({\rm Sym}  (O_{ab} )  | ~\Big|~  1\le a\le  b \le r, \,  x\in\set X\Big)$ are linearly independent.       
\end{cor}

The proof follows from lemmas \ref{lem:hermitianmatrix},  \ref{lem:complexindep} and \ref{lem:realindep} in Appendix \ref{app:extrememapping}.  

Proposition \ref{prop:povm} also implies a characterisation of the extreme points of various sets of quantum instruments $(\map C_x)_{x\in\set X}$ with $\map C_x :  L(\spc H_{\rm in}) \to L(\spc H_{\rm out}), \forall x\in\set X$.      
 Denoting by $\map C_x  (\rho)   =  \sum_{a=1}^{r_x}  \,   C_a^{(x)} \rho  C_a^{(x) \dag}$ a minimal Kraus representation of the map $\map C_x$, and considering the complex case,  we have the following characterisations:
  \begin{itemize}
\item {\em Extremal quantum instruments \cite{pellonpaa2012quantum}.}    A quantum instrument $(\map C_x)_{x\in\set X}$ satisfies the condition that the map $\map C_{\set X}:  =\sum_{x\in \set X} \map C_x$ is trace-preserving. The extremality condition in   Proposition \ref{prop:povm}  reduces to the requirement that   the operators  $  \left(C^{(x)\,\dag}_a  C^{(x)}_b\right)_{1\le a\le r,  1\le b\le r \, ,x\in\set X}$ be linearly independent.  
\item {\em Extremal bistochastic instruments.}     A bistochastic instrument \cite{chiribella2021symmetries}  satisfies the condition that the map $\map C_{\set X}$ is both trace-preserving and identity-preserving.  The extremality condition in   Proposition \ref{prop:povm}  reduces to the requirement that   the operators $  \left(C^{(x)}_a  C^{(x)\, \dag}_b  \oplus  C^{(x)\, \dag}_a C^{(x)}_b\right)_{1\le a\le r,  1\le b\le r , \, x\in\set X }$ be linearly independent.  
\item {\em Extremal  Gibbs-preserving  instruments.}  We can define a Gibbs-preservig instrument as an instrument that satisfies the condition that the map $\map C_{\set X}$ is both trace-preserving and Gibbs-preserving, that is $\map C_{\set X}  (\rho_{\rm in})  =  \rho_{\rm out}$ for some fixed density matrices $\rho_{\rm in}$ and $\rho_{\rm out}$, interpreted as Gibbs states of the input and output system, respectively.    The extremality condition in   Proposition \ref{prop:povm}  reduces to the requirement that   the operators $  \left(C^{(x)}_a   \rho_{\rm in} C^{(x)\, \dag}_b \oplus  C^{(x)  \, \dag}_a C^{(x)}_b\right)_{1\le a\le r,  1\le b\le r , \, x\in\set X }$ be linearly independent.    
\item {\em Extremal  instruments that map a given input POVM into a given output POVM.}     It corresponds to the case $k=0$,  $A_j= P_j$,   $B_j=  Q_j$ for two given POVMs $(P_j)_{j=1}^n$ and $(Q_j)_{j=1}^n$    In this case,  Eq. (\ref{extrememapping}) reduces to the requirement that   the operators  $  (\bigoplus_{j=1}^n  \, C_a^\dag Q_j  C_b )_{1\le a\le r,  1\le b\le r}$ be linearly independent.  
\end{itemize} 
Similar characterisations can be obtained in the real case, replacing the operators in the above sets with their symmetric version and restricting the range of the indices $a$ and $b$ to $1\le a\le b\le r$.

\section*{Acknowledgments}
This work has been supported by the Hong Kong Research Grant Council through  grant no.\ 17307520 and through the Senior Research Fellowship Scheme SRFS2021-7S02, and by the John Templeton Foundation through grant  62312, The Quantum Information Structure of Spacetime (qiss.fr).   Research at the Perimeter Institute is supported by the Government of Canada through the Department of Innovation, Science and Economic Development Canada and by the Province of Ontario through the Ministry of Research, Innovation and Science. The  opinions expressed in this publication are those of the author and do not necessarily reflect the views of the John Templeton Foundation.

\appendix  

\section{Proof of Proposition \ref{prop:extrememaxent}}\label{app:maxent}

 Every operator  $\rho  \in  \Herm_+  (\spc H_1\otimes \spc H_2)$ can be diagonalised as  
   \begin{align}\label{diagrho}
   \rho   =  \sum_{a=1}^r  |\Psi_a\>\<\Psi_a|  \, , \qquad{\rm with}  \qquad |\Psi_a\>\not  =  0 ~  \forall  a  \quad {\rm and} \quad  \<\Psi_a|  \Psi_b\>  =  0  \, , \forall a\not =  b  \, .   
   \end{align}
In addition,  one can always find linear operators $   F_a:    L(\spc H_1) \to L(\spc H_2)$ such that
\begin{align}\label{doubleket}
 |\Psi_a\>  =  (  I_{1}  \otimes  F_a)   \,  |\Phi\>  \qquad {\rm with} \qquad |\Phi\>  :  =  \frac 1{\sqrt{d_1}} \,\sum_{k=1}^{d_{1}}  \,  |k\>\otimes |k\> \in \spc H_{1} \otimes \spc H_{1} \, . 
 \end{align} 
 With this notation, the marginals of the operator $\rho$ are  
 \begin{align}\label{marginals}
 \Tr_1  [  \rho]   =  \frac 1{d_1} \sum_{a=1}^r   \,  F_aF_a^\dag  \qquad{\rm and} \qquad  \Tr_2  [  \rho]   =   \frac 1 {d_1}  \sum_{a=1}^r   \,  F_a^T \overline F_a \, ,
 \end{align}
 where  $F_a^T$ and $\overline F_a$  are the transpose and the complex conjugate of $F_a$  with respect to the standard bases of $ \spc H_{1}$ and $ \spc H_{2}$, respectively.

Now, suppose that $\rho$ is an extreme point of ${\sf S}  \left(\frac I2,  \frac I2\right)$.  By Eqs. (\ref{cmarg}) and (\ref{rmarg})  we know that $r$  is at most $2$, both in the real and in the complex case.       Hence, Eq. (\ref{marginals}) implies   $F_1F_1^\dag  +  F_2  F_2^\dag =  I$ and $F_1^T  \overline F_1  +  F_2^T \overline F_2 =  I$ or equivalently, 
\begin{align}\label{bastaa}
F_1F_1^\dag  +  F_2  F_2^\dag =    I   \qquad {\rm and}  \qquad  F_1^\dag  F_1  +  F_2^\dag  F_2 =  I\, ,
\end{align}

  The first equation implies that    $F_1F_1^\dag  $ and $ F_2  F_2^\dag$ are diagonal in the same basis.    Hence,  $F_1  =  \sqrt{P}  \, U_1$  and $F_2  =  \sqrt{I-P}     \,U_2$  for some unitary operators $U_1$ and $ U_2$ and some positive contraction $0\le P\le I$.    The second equation in Eq. (\ref{bastaa}) implies that $U_1^\dag P \,U_1  +  U_2^\dag (I- P)  U_2  =  I $, which implies $U_1^\dag   P   U_1  =  U_2^\dag    P  U_2$, or equivalently $[P,  U_1 U_2^\dag]  =  0$.  
 Now, using Eqs.  (\ref{diagrho}) and (\ref{doubleket}), the state $\rho$ can be rewritten as  
  \begin{align}\label{rhog}
  \rho     =   (\map I_1 \otimes \map G  \map U_2) \, (|\Phi\>\<\Phi|)   =  ({\map U}_2^T \otimes \map G)   \, (|\Phi\>\<\Phi|)   \, ,
  \end{align} 
  where $\map U_2,  \map U_2^T$, and $\map G$ are the completely positive maps defined by $\map U_2 (A)  :=   U_2 A U_2^\dag$,   ${\map U}_2^T (A)  :  =   U_2^T  A  \overline U_2$, and  $\map G  (A):  G_1\rho G_1^\dag +  G_2A G_2^\dag$, with
  \begin{align}
  G_1  :  =   F_1   U_2^\dag     =   \sqrt P  \,   U_1U_2^\dag   /\sqrt 2  \qquad  {\rm and}   \qquad G_2:  =  F_2  U_2^\dag  =  \sqrt{I-P}/\sqrt 2  \, . 
  \end{align}
 The operators $G_1$ and $G_2$ are normal, and therefore unitarily diagonalizable.  Moreover, they commute with each other, and therefore they are diagonal in the same orthonormal basis $\{|\psi_1\>, |\psi_2\>\}$.   Let us write $G_1  =  \alpha  \, |\psi_1\>\<\psi_1|  +  \, \beta  |\psi_2\>\<\psi_2|$ and  $G_2  =  \gamma  \, |\psi_1\>\<\psi_1|  +  \, \delta  |\psi_2\>\<\psi_2|$, for suitable numbers $\alpha,  \beta,\gamma,  \delta \in \K$ satisfying the condition $|\alpha|^2  +  |\gamma|^2  =  |\beta|^2  +   |\delta|^2    =1$ (required by the condition $G_1G_1^\dag  +  G_2  G_2^\dag=  F_1F_1^\dag  +  F_2  F_2^\dag =  I$).  Using the decomposition of $G_1$ and $G_2$, the map $\map G$ can be rewritten as 
 \begin{align}
 \map G  (A)   =  \sum_{a,b=1}^2 \,   \xi_{ab}  \,  \<\psi_a|  A  |\psi_b\>  \,  |\psi_a\>\<\psi_b|  \,  , \qquad  \xi  =  \begin{pmatrix}
 1  &   \alpha\overline \beta   +  \gamma \overline \delta  \\
 \overline \alpha \beta +  \overline \gamma \delta  & 1  \,.
 \end{pmatrix}
 \end{align}  
  The matrix $\xi$ is a  $2\times 2$ correlation matrix.  Hence,    $\map G$ is a Hadamard channel (as defined in subsection \ref{subsec:hadamard}).  Corollary \ref{cor:unitaryextr} them implies that $\map G$ is a convex combination of unitary channels.  Hence, Eq. (\ref{rhog}) implies that $\rho$ is a mixture of maximally entangled states. \qed

\section{Proof of Proposition \ref{prop:extrememapping}}\label{app:extrememapping}

The proof of Proposition \ref{prop:extrememapping} uses the following lemmas:  
  \begin{lemma}\label{lem:hermitianmatrix}
Let $\spc H_{\rm in}$ and $\spc H_{\rm out}$ be Hilbert spaces over $\K$, with $\K  = \R$ or $\K  =  \C$, and let $C=  \sum_{a=1}^r   |\Gamma_a\>\<\Gamma_a|$ be an element of ${\sf S}_{\rm map}$, with linearly independent vectors $\{|\Gamma_a\>\}_{a=1}^r$, and, for every $a\in  \{1,\dots,  r\}$, let $C_a:  \spc H_{\rm in}\to \spc H_{\rm out}$ be the operator such that $|\Gamma_a\>    =    (I_{\rm in}\otimes  C_a) \,  |\Phi\> \,,$ with $|\Phi\>:  =  \sum_{l=1}^{d_{\rm in}} \,  |l\>\otimes |l\>$.   The operator $  C$   is an extreme point of ${\sf S}_{\rm map}$ if and only if the implication
  \begin{align}
  \nonumber   & \sum_{a,b }  \,  x_{ab}  \,     \left[ \left( \bigoplus_{j=1}^k C_a    A_j   C_b^\dag \right)\oplus\left(  \bigoplus_{j=k+1}^n  C_a^T   B_j^T\,   \overline C_b\right) \right]     =   0  \\
 & \qquad   \qquad    \qquad   \qquad \qquad  \qquad   \qquad   \Longrightarrow  \quad  x_{ab}  =  0  ~ \forall a, b \in  \{1,\dots,  r\}  \label{xindep}
    \end{align} 
  holds for every Hermitian  $r\times r$ matrix $(x_{ab}) \in  M_r (\K)$.
  \end{lemma} 
 \Proof   
 Let    $ \map M  :     L(\spc H_{\rm in} \otimes \spc H_{\rm out})   \to  L(\spc H_{\rm out}) \oplus   L(\spc H_{\rm in})$ be the linear map defined by 
 \begin{align}\label{migoplus}  
\map M  (A)   := \left(  \bigoplus_{j=1}^k   \Tr_{\rm in}   [  (A_j\otimes  I_{\rm out})   \,A]  \right)  \oplus  \left(  \bigoplus_{j=k+1}^n    \Tr_{\rm out}   [  ( I_{\rm in } \otimes B_j)  \,A] \right)        
 \end{align}   
 By Theorem \ref{theo:extreme1}, extremality of $C$ is equivalent to the injectivity  of $\map M$ on $\Herm (\Supp (C))$. 
Now,    every operator  $H  \in \Herm  (\Supp (C))$  can  be written as   $H = \sum_{a,b=1}^r  \, x_{ab}  \, |\Gamma_a\>\<\Gamma_b|$,  where   $(x_{ab}) \in M_r (\K)$ is a Hermitian matrix. 
  Using the relation $|\Gamma_a\>    =    (I_{\rm in}\otimes  C_a) \,  |\Phi\> $,  we  can transform Eq. (\ref{migoplus}) into
  \begin{align}
  \map M(H)   = \sum_{a,b }  \,  x_{ab}  \,     \left[ \left( \bigoplus_{j=1}^k C_a    A_j   C_b^\dag \right)\oplus\left(  \bigoplus_{j=k+1}^n  C_a^T   B_j^T\,   \overline C_b\right) \right]    \, .
  \end{align} 
Since $(x_{mn})  \in  M_r (\K)$ is an arbitrary Hermitian matrix, the  injectivity of the map $\map M$ is equivalent to Eq. (\ref{xindep}).\qed 

\begin{lemma}\label{lem:complexindep}
Let $r$ be an integer, $\spc H$ be a complex Hilbert space, and  let $\{O_{ab} \}_{1\le a \le r\,  1\le b\le r} \subset  L(\spc H)$   be a set of operators satisfying the condition  $O_{ab}^\dag  =  O_{ba}$, $\forall a,  b$.   The  following conditions are equivalent: 
\begin{enumerate}
\item   for every Hermitian matrix $(x_{ab})  \in  M_r (\C)$, $\sum_{a,b }  \,  x_{ab}  \,    O_{ab}  =   0$ implies $x_{ab}  =  0  ~ \forall a, b \in  \{1,\dots,  r\}$,
\item the operators $(   O_{ab})_{1\le a\le r, \, 1\le b\le r}$ are linearly independent over $\C$.  
\end{enumerate}  
\end{lemma} 

\Proof  The implication $2\Rightarrow 1$ is trivial.  Hence, we only need to prove the implication $1\Rightarrow 2$.    Suppose that $\sum_{a,b}   c_{ab}  \,   O_{ab}   =  0$ for some complex coefficients $(c_{ab})_{a,b}$.   Taking the adjoint on both sides of the equation, we obtain  $\sum_{a,b}   \overline c_{ab}  \,   O^\dag_{ab}   =  0$, or equivalently $\sum_{a,b}   \overline c_{ba}  \,   O_{ab}   =  0$.  Hence, we have $\sum_{a,b}    (c_{ab} +  \overline c_{ba}  ) \,   O_{ab}   =  0$ and $\sum_{a,b}   i (c_{ab} -  \overline c_{ba}  ) \,   O_{ab}   =  0$.   Since the matrices   $ (c_{ab} +  \overline c_{ba}  )$ and $i   (c_{ab} -  \overline c_{ba}  )$ are both Hermitian, Condition 1 implies that $ (c_{ab} +  \overline c_{ba}  )$ and $i   (c_{ab} -  \overline c_{ba}  )$ are both zero, or equivalently  $c_{ab}  =  0 \, ,\forall a,b$. Hence, the operators $(  O_{ab})_{a,b}$ are linearly independent.  \qed

\begin{lemma}\label{lem:realindep}
Let $r$ be an integer, $\spc H$ be a real Hilbert space, and  let $\{O_{ab} \}_{1\le a \le r\,  1\le b\le r} \subset  L(\spc H)$   be a set of operators satisfying the condition  $O_{ab}^T  =  O_{ba}$, $\forall a,  b$.   The  following conditions are equivalent: 
\begin{enumerate}
\item   for every real symmetric matrix $(x_{ab})  \in  M_r (\R)$, $\sum_{a ,b}  \,  x_{ab}  \,    O_{ab}  =   0$ implies $x_{ab}  =  0  ~ \forall a, b \in  \{1,\dots,  r\}$,
\item the operators $(  {\rm Sym} (O_{ab}))_{1\le a  \le  b\le r}$ are linearly independent over $\R$.  
\end{enumerate}  
\end{lemma} 

\Proof  $1\Rightarrow 2$.    Suppose that $\sum_{1\le a \le b\le r}   c_{ab}  \,   {\rm Sym}  (O_{ab} )  =  0$ for some real coefficients $(c_{ab})_{a,b}$.   This condition implies  $\sum_{a  =1}^r \sum_{b=1}^r   x_{ab}  \, O_{ab}   =  0$, where $(x_{ab})$ is the real symmetric matrix with $x_{ab}:  = c_{ab} \,  \forall 1\le a \le  b\le  r$.   Since the matrix   $ (x_{ab} )$ is symmetric, Condition 1 implies that all its entries  are zero, or equivalently  $c_{ab}  =  0 \, ,\forall a,b$. Hence, the operators $(  {\rm Sym}  (O_{ab}) )_{1\le a\le b \le r}$ are linearly independent. 

 $2\Rightarrow 1$. Suppose that $\sum_{a=1}^r\sum_{b=1}^r  \,  x_{ab}  \,    O_{ab}  =   0$  for some real symmetric matrix $(x_{ab})$.  Hence,  $\sum_{1\le a\le b\le r}    x_{ab}  \,  {\rm Sym}   (O_{ab}) =0$. Since the operators $(  {\rm Sym}  (O_{ab}) )_{1\le a\le b \le r}$ are linearly independent, one must have  $x_{ab}   = 0  \, \forall a,b$. 
 \qed

{\bf Proof of Proposition \ref{prop:extrememapping}.}       For $1\le a\le r$ and  $1\le b\le r$,  
   define the operators $O_{ab}  :   =   \left( \bigoplus_{j=1}^k C_a    A_j   C_b^\dag \right)\oplus\left(  \bigoplus_{j=k+1}^n  C_a^T   B_j^T\,   \overline C_b\right)$.    Note that one has $O_{ab}^\dag  =  O_{ba} \,  ,\forall a,\forall b$.   Lemma \ref{lem:hermitianmatrix},  the operator $C $ is extremal if and only if, for every Hermitian matrix $(x_{ab}) \in  M_r (\K)$,   the condition $\sum_{a,b}   \,  x_{ab} \,  O_{ab} =0$ implies $x_{ab}  = 0\, ,  \forall a, b$.  In the complex case,  Lemma \ref{lem:complexindep} implies that the operators $(O_{ab})_{1\le a \le r,  1\le b\le r}$ are linearly independent over $\C$.   Note that the linear independence of the operators $(O_{ab})_{1\le a \le r,  1\le b\le r}$  is equivalent to the linear independence of the operators    $\left( \bigoplus_{j=1}^k C_a    A_j   C_b^\dag \right)\oplus\left(  \bigoplus_{j=k+1}^n  C_a   B_j\,   C_b\right)\, , 1\le a\le r,  1\le b\le r$.      In the real case,  Lemma \ref{lem:realindep} implies that the operators $({\rm Sym}  (O_{ab}))_{1\le a \le b\le r}$ are linearly independent over $\R$.   Note that in the real case $O_{ab}    =   \left( \bigoplus_{j=1}^k C_a    A_j   C_b^T \right)\oplus\left(  \bigoplus_{j=k+1}^n  C_a^T   B_j\,   C_b\right)$.  \qed

\bibliographystyle{ws-ijqi}
\bibliography{extremality}
\end{document}